%% file: ms_rev1.tex
\documentclass{ar-1col-S2O}

\usepackage[comma]{natbib}
\usepackage{url}

\input{ads_macros.tex}

\usepackage{amssymb,amsmath}
\usepackage[normalem]{ulem}
\usepackage[colorlinks]{hyperref}
\usepackage[capitalize]{cleveref}
\usepackage{csquotes}
\usepackage{bm}

\newcommand{\hunit}{\text{km/s/Mpc}}
\newcommand{\inversempc}{\ensuremath{\text{Mpc}^{-1}}}

\jname{Xxxx. Xxx. Xxx. Xxx.}
\jvol{AA}
\jyear{YYYY}
\doi{10.1146/((please add article doi))}

\begin{document}

\markboth{Verde et al.}{A tale of many $H_0$}

\title{A tale of many $H_0$}
\author{Licia Verde,$^{1,2}$ Nils Sch\"oneberg,$^1$ and H\'ector Gil-Mar\'in$^1$
\affil{$^1$ ICC-UB Institut de Ci\`encies del Cosmos, University of Barcelona,\\
Mart\'i i Franqu\`es, 1, E-08028 Barcelona, Spain, email: liciaverde@icc.ub.edu}
\affil{$^2$ ICREA, Pg. Llu\'is Companys 23, Barcelona, E-08010, Spain}
}

\begin{abstract}

\hangindent=.3cm$\bullet$
 The Hubble parameter $H_0$,  is not a univocally-defined,  quantity: it  relates redshifts to distances in the near Universe, but is also a key  parameter of the  $\Lambda$CDM standard cosmological model. As such, $H_0$ affects several physical processes at different cosmic epochs, and multiple observables.  We have counted more than a dozen $H_0$'s which are expected to  agree if a) there are no significant systematics in the data and their interpretation and b) the adopted cosmological model is correct. 

 \hangindent=.3cm$\bullet$ With few exceptions (proverbially confirming the rule) these determinations do not agree at high statistical significance; their values cluster around two camps: the low (68 km/s/Mpc) and high (73 km/s/Mpc) camp. It appears to be a matter of anchors: the shape of the Universe expansion history agrees with the model, it is  the normalizations that disagree.
 
 \hangindent=.3cm$\bullet$  Beyond systematics in the data/analysis, if   the model is incorrect there are only two viable ways to “fix” it: by changing the early time ($z\gtrsim 1100$) physics and thus the early time normalization, or  by a global modification,  possibly touching the model's fundamental assumptions (e.g., homogeneity, isotropy, gravity).
 None of these three options has the consensus of the community.
 
  \hangindent=.3cm$\bullet$ The research community has been actively looking for deviations from $\Lambda$CDM for two decades; the one we might have found makes us wish we could put the genie back in the bottle. 

\end{abstract}
\begin{keywords}
cosmology, cosmological parameters, Hubble parameter, cosmological distances, cosmic microwave background, large-scale structure
\end{keywords}

\maketitle

%
\tableofcontents 
%
\section{INTRODUCTION}
The current expansion rate of the Universe is a fundamental quantity in cosmology.
It is represented by the Hubble constant $H_0\equiv 100 h$\,\hunit, and it readily allows for the conversion of  recession velocities (which are relatively easy to measure via redshifts) to distances, which are notoriously difficult to measure (but great progress has been made in the last 20 years). The subscript $0$ indicates that it is a local quantity measured at the present day, at redshift $z=0$.   
However, it is customary to also relate the expansion rate of the Universe at higher redshifts to this present-day value via $H(z) \equiv H_0 E(z)$, since many astrophysical probes only measure the uncalibrated (unnormalized) expansion rates $E(z)$. This formulation, and the appearance of $H$ in the Friedmann equations,  makes it evident that $H_0$ is also a global quantity, and $h$ therefore becomes a universal yardstick.  So much so that the \enquote{little h} appears everywhere in astrophysics and cosmology, not only in background quantities but in any measurement or constraint where an assumption about distances must be made \citep{Croton2013}. \begin{marginnote}[]\entry{Hubble constant, $H_0$}{The current expansion rate of the Universe.  It is also a key parameter of any cosmological model.}
\end{marginnote}

It is therefore unsurprising that $H_0$ (and $h$) is one of the key parameters of cosmology.

The standard model of cosmology is the $\Lambda$CDM model,%
\begin{marginnote}[]\entry{$\Lambda$CDM}{The standard model of cosmology, a spatially flat, cosmological constant-dominated Universe, with cold dark matter and baryons as the major components that cluster under gravity.}
\end{marginnote}%
a spatially flat, cosmological constant-dominated Universe, with cold dark matter and baryons as the major components that cluster under gravity, and three neutrino species. Despite having its two major components (dark matter and dark energy) being poorly understood and probably somewhat {\it ad hoc},  this model has enjoyed spectacular success over the past three decades:  an avalanche of data (for example \cite{wmap9, PlanckParams2020, Alam2021}, etc., and references therein) confirmed  the concordance of the model with a wide suite of detailed observations of the Universe across most of its evolution (from recombination at $z\sim 1100$ to the present day). The parameters of the model are now constrained with exquisite precision, sometimes better than at the percent level. 

Observations of the Cosmic Microwave Background (CMB)%
\begin{marginnote}[]
\entry{CMB}{Cosmic Microwave Background. The first light that could ever travel freely throughout the Universe.}
\end{marginnote}%
have been pivotal in establishing the model and constraining its parameters. In particular, within this model, from CMB observations alone, the Hubble constant is constrained at the 0.7\% level (hereafter, unless otherwise stated,  error-bars correspond to $1\sigma$, 68\% confidence level): $H_0=67.4 \pm  0.5\,\hunit$ \citep{PlanckParams2020}. Of course, observations of the early Universe are not directly sensitive to the present-day expansion rate nor measure the distance-redshift relation at $z=0$: this is an indirect (and thus model-dependent) constraint. The CMB is extremely sensitive to the expansion rate of the Universe at and around the time of recombination; within the standard $\Lambda$CDM model the constraints on recombination are \enquote{translated} (with little wiggle room) into current day quantities. This model quite rigidly maps the early-time Universe into the late-time one and vice-versa.

Until the mid 2010s -- despite progressively shrinking uncertainties -- local measurements of the Hubble constant from the distance-redshift relation that are model-independent, agreed well with the model-dependent indirect constraints coming from the early-Universe. 
\begin{figure}
\begin{minipage}[c]{0.6\linewidth}
\centering
   \includegraphics[width=2.7in,trim = 0.2cm 0cm 0cm 0cm, clip]{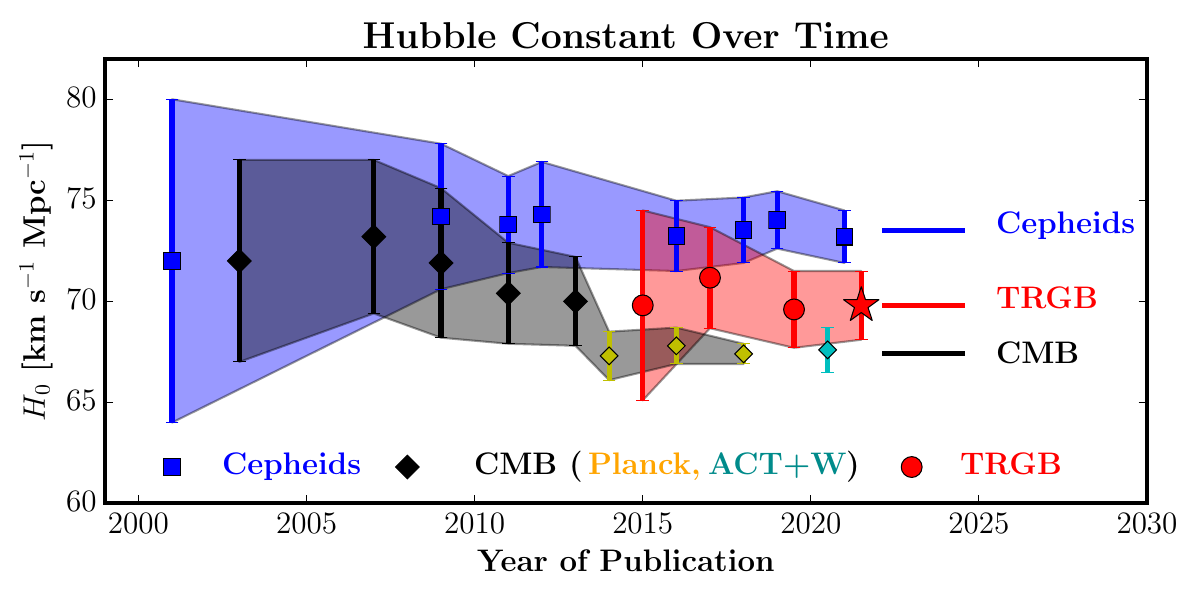}
   \end{minipage}
  \begin{minipage}[c]{0.4\linewidth}
\centering 
    \includegraphics[width=1.9in, trim = 4cm 10.5cm 4cm 10.5cm, clip]{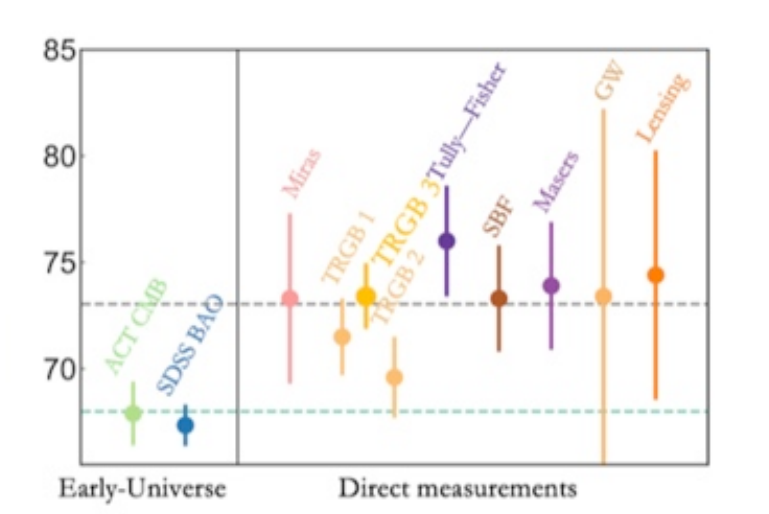}
    \end{minipage}
    \caption{Left panel: summary of the evolution of the Hubble constant constraints over the past two decades: measurements based on Cepheid variables (blue, squares), inference based on CMB observations (gray)  WMAP (black diamonds) Planck (yellow diamonds), and ACT+WMAP (cyan diamonds). In red are measurements based on the tip of the red giant branch (see \cref{sec:TRGB}). Note how the agreement  between local and \enquote{global} $H_0$ determinations degrades around 2013 with the advent of the Planck data (reproduced from Fig.~11 of \cite{Freedman21}). Each data point is a representative value of the (sometimes several) publications in that year, hence some data points have broader \enquote{constituencies} than others, as we hope it will be clear from the rest of this review. The left panel is a zoom-in on recent developments circa 2022-2023,  which acronyms will become clearer in the following chapters of this review, the dashed lines guide the eye indicating the central values of the "Cepheids" and "CMB" regions of the left panel; courtesy of Anowar Shajib.}
    \label{fig:H-year}
\end{figure}
This represented one of the triumphs of the standard model of cosmology, leading to  the allegory of \enquote{threading a needle from the other side of the Universe} (A. Riess 2018, MIAPP workshop). This all changed around 2013, as shown in 
\cref{fig:H-year}, reproduced  in part from from Fig.~11 of~\cite{Freedman21}, which reports the state-of-the-art $H_0$ measurements as a function of publication year over the past two decades.

Local measurements%
\begin{marginnote}[]
\entry{Local $H_0$}{Measurements of the Hubble constant that rely on measuring the distance-redshift relation.}
\entry{Distance indicators}{Standard objects, whose intrinsic properties can easily be inferred and compared to the properties that we observe.}
\end{marginnote}%
of the Hubble constant mostly rely on building a cosmic distance ladder with several staggered distance indicators, where each step of the ladder is calibrated  on the previous step. Distance indicators are standard objects, whose  intrinsic properties can easily be inferred and compared to the properties that we observe hence yielding a measurement of  their distance.

These local measurements tend to cluster around $H_0$ values higher (by about 7\%) than those inferred from the distant universe. The figure also shows that the most precise local constraint, based on the Cepheid-variables, reaches a $\sim$1\% level precision.

The fact that $h$ (as the global parameter of a cosmological model) and the proportionality of low $z$ distance-redshift relation are identified to be the same quantity, requires significant model assumptions such as homogeneity, isotropy, and all other assumptions on which the adopted model is built upon. For the standard cosmological model these include 
general relativity, the existence of cold dark matter, a cosmological constant, etc. Agreement of this model with observations among a wide range of redshifts represents a validation of the combination of all these assumptions. Conversely, a mis-match or lack of agreement seeds doubts on the reliability of the standard model of cosmology. 

Between 2013 and 2019 (e.g., \cite{Planck2013, tension, bernal_trouble_2016,VTR_2019} and references therein) this mismatch became increasingly evident, to the point that the term \enquote{Hubble tension} was coined.%
\begin{marginnote}
\entry{Hubble tension}{The mismatch between local $H_0$ determinations and the inferred cosmological parameter value 
in the $\Lambda$CDM model.}
\end{marginnote}

Today, ten years after the first tentative indications of a tension, it is interesting to look back and note how the reactions of the community to this (perceived) mismatch evolved. 

Throughout, significant efforts went into checking for systematics in all related measurements. As part of this effort, there was renewed interest in looking for alternative or complementary probes and experiments. As of today, the hunt for systematic errors can be said to have improved the reproducibility of findings, and robustness of all measurements, and ultimately strengthened the evidence of a tension.

It is also fair to say that the developments of the last decade have changed the expectations and {\it modus operandi} of a big part of the community. The community now expects results to be reproducible, hence the data and key software to be publicly available in such a way that a practitioner not involved in the original analysis could still retrace and reproduce all important steps and findings. While research areas such as the CMB and large-scale structure made this transition to \enquote{open science} about two decades ago, this was not the case for other areas of extra-galactic astronomy, but this os now changing. 

Ten years on and despite all this effort, the tension still remains. An explanation based on very simple systematic errors would likely have emerged, so invoking systematics seems an increasingly unlikely  successful explanation. The stakes are high:
if the Hubble tension represents a symptom of a systemic failure of the standard cosmological model or an indication that the model is incorrect or needs an extension, it is not clear which aspects of the model or what model assumptions may need to be modified.   

In this light, a Hubble tension is not \enquote{just} a Hubble tension. Any \enquote{fix} by changing the model will typically show up in some (other) observations, possibly unrelated to $H_0$. In the era of precision cosmology, this both severely restricts attempts of model building, but also allows for many proposals to be rapidly falsified.

The quest for alternative probes and measurements has marked different routes to $H_0$ and  produced several different $h$'s: these approaches can be said to  measure the same quantity  only under specific model assumptions: their agreement (or lack of it) can be used to test  specific ingredients of the model itself.

The goal of  this review is to summarize the current status of the various determinations of the Hubble constant $H_0$, offering enough background explanation to understand the strengths and limitations of all the main constraints, and most importantly to explore their synergies in view of testing the standard cosmological model. To this aim, selected determinations throughout the text are marked by a code in [], and used as  input to  the summary figure, \cref{fig:summary}. 
We refer to other excellent reviews which have appeared recently \cite{Kamion_Riess22, DiValentino:2021izs, diValentinoIntertwined} and \cite{Wendy23} for complementary perspectives. The first one focuses on a specific  promising theoretical solution i.e.,  a modification of the standard cosmological model. The second  and third focus on detailing the theoretical efforts in searching for a solution; the fourth is an historical and critical review of the local $H_0$ determinations.
 For this reason in this review we will not discuss  at length specific  modifications to the cosmological model (or physics) needed to resolve the tension.  We aim instead at reviewing the guardrails (see Sec.\ref{ssec:guardrails})  on the road to a  theoretical solution that current observations can provide. Any successful modification of the cosmological model or any  proposed new physics that aim at resolving the Hubble tension has to conform to the
measurements of all available observations within the reported uncertainties and with the caveats discussed here.

\section{DIRECT DISTANCE LADDER(S)}\label{sec:distance}

One of the most direct methods to measure the Hubble constant is through the measurement of the Hubble flow\begin{marginnote}[]
\entry{Hubble flow}{The expected observed recession of distant objects caused purely by the expansion of the universe.}
\end{marginnote}%
in our cosmic neighbourhood. To perform this measurement both the distances and recession velocities (typically characterized through the redshift) of a large number of distant objects in the Hubble flow is required. While the measurement of the redshift is often more or less straightforward for these objects, it is the estimation of their distances that is usually the more difficult problem.

While for relatively nearby objects well-known methods such as parallax measurements are sufficient, more distant objects are not yet within the reach of current resolution. One of the most straightforward ways to gauge the distance to such a faraway cosmological object is to compare the inherent properties of the object to those that are observed from a distant observer such as ourselves. 
The problem of estimating a distance is then reduced to the estimation of the intrinsic properties of the object, most typically the emitted brightness.

To do so a vast collection of calibration techniques has been developed in the past. These techniques rely on the cosmological principle and the derived assertion that faraway objects should behave like nearby ones when studied in the same galactic environment, given fair or representative samples of each. After correcting for environmental influences, such objects of known brightness (also known as standard candles) or of known size (standard rulers) are a crucial pillar of Hubble constant measurements.

The most abundant and precise standard candles\footnote{We will use \enquote{standardizable} and  enquote{standard} object/candle interchangeably. The used objects  always have to be corrected/standardized (e.g., environment, metallicity) to be used as standard candles.} to measure the Hubble flow even up to high redshift are the bright supernovae of type Ia (SNeIa).\begin{marginnote}[]
\entry{SNIa}{Type Ia Supernova. 
Standard(-izable) candle.}
\end{marginnote}%
However, crucially, there is a very limited number of nearby supernovae:  an intermediate calibrator is required.
This is the basis of the distance ladder, which at each rung calibrates a more distant set of standardizable objects. The exact objects used can vary from analysis to analysis. 
A few of the most common choices are discussed below.

The first rung of the distance ladder typically consists of a geometric measurement and calibration of standardizable objects in our galaxy or nearby galaxies (like the Large Magellanic Cloud, the Andromeda galaxy, or M106). The galaxies in the first rung are typically called the {\it anchors},\begin{marginnote}[]
\entry{Anchors}{The nearby galaxies in the first rung of the cosmic distance ladder that set the overall absolute scale for the calibration of distance indicators.}
\end{marginnote}%
since they set the overall absolute scale of the intrinsic property used for calibration, and thus anchor the Hubble constant measurement to local objects. The standardizable objects (calibrators), which are calibrated based on the anchor distances, are then in turn used within the second rung of the distance ladder to calibrate faraway standardizable objects (such as SNeIa), which can be finally used in the third rung of the distance ladder to infer the Hubble flow and correspondingly the Hubble constant.  
\subsection{Building the standard direct distance ladder}
\begin{figure}
    \centering
    \includegraphics[width=0.85\textwidth, angle=-90, trim = 0cm 2.5cm 0cm 0cm, clip]{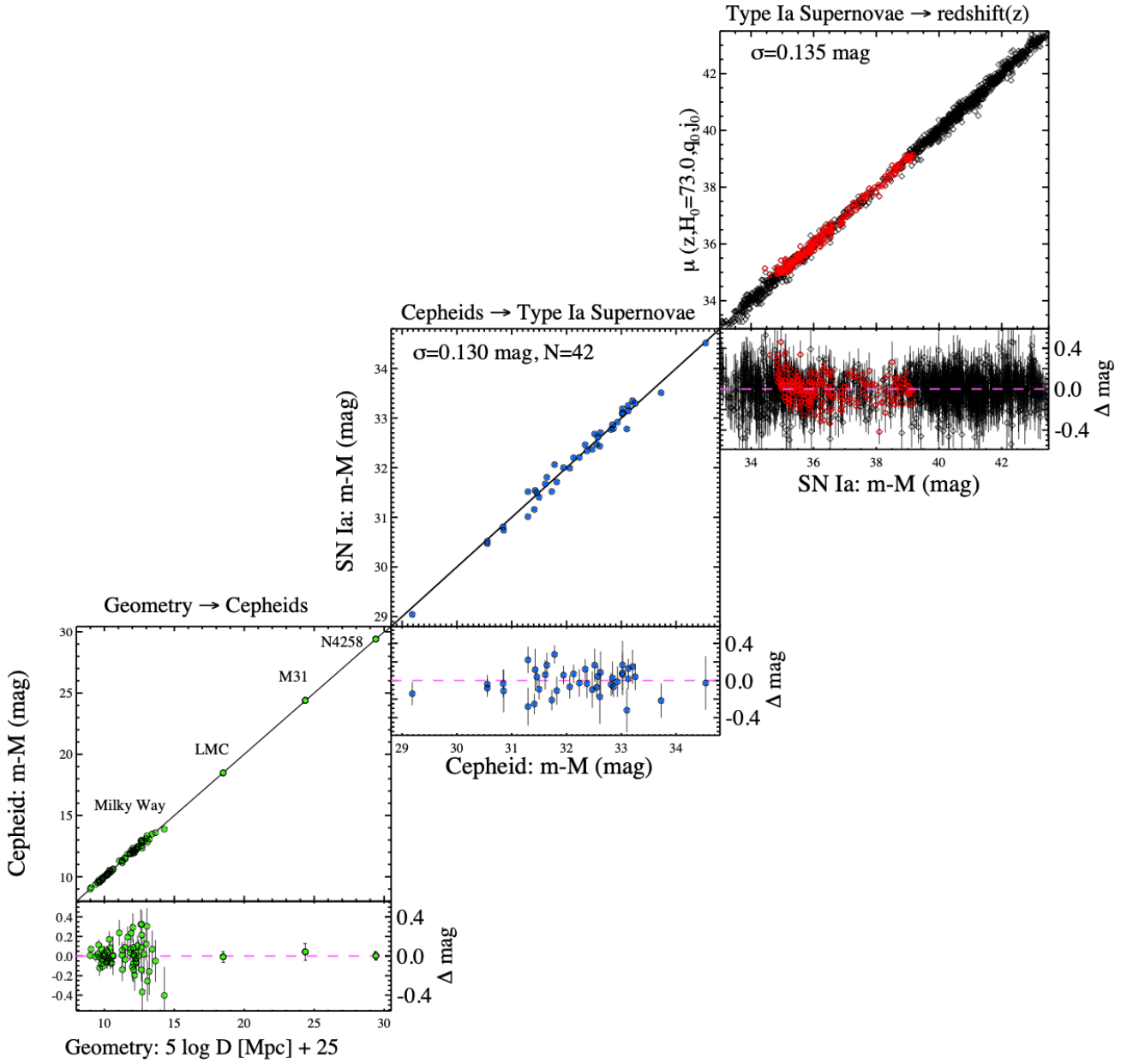}
    \caption{Distance ladder based on Cepheids. Figure reproduced from \cite{Riess2022b}. This is a three rung ladder with two key calibration steps. The bottom left panel illustrated the geometric distance to Cepheids. This calibrates the Cepheid distances based on the period-luminosity relation.   The middle square shows the calibration of SNeIa to Cepheid distances in SNeIa host galaxies. The top right square shows Supernovae in the Hubble flow. Those  satisfying the redshift cuts (in red)  that, once calibrated, yield the Hubble constant.}
    \label{fig:ladder}
\end{figure}

While many options exist for constructing a distance ladder, currently the most precise constraints on the Hubble constant rely on  the vast abundance, high luminosity and the very predictable nature of Cepheid variables as calibrators and on supernovae of type Ia (SNeIa) as cosmological distance indicators. The standard, \enquote{classic} cosmic distance ladder involves three rungs.  These are described below and illustrated in \cref{fig:ladder}.

\paragraph{First rung of the ladder}
Cepheids\begin{marginnote}[]
\entry{Cepheids}{Pulsating stars with a tight period-luminosity relation.  Excellent standard candles and distance indicators.}
\end{marginnote}%
are pulsating stars whose $\kappa$-opacity mechanism drives a tight period-luminosity relation, first observed by Henrietta Swan Leavitt  \citep{Leavitt}. 
The most important systematic factors influencing the accuracy of the Cepheid-based determination are dust extinction (affecting the apparent luminosity) and the correlation with metallicity (which affects the intrinsic luminosity). 

The dependence on dust extinction is usually modeled through the use of Wesenheit magnitudes (based on a given reddening law of dust), while the metallicity is measured through abundance ratios of strong emission lines in HII regions. In total, the slope of the period-luminosity relation can be measured to permille level accuracy, making Cepheids excellent calibrators. It then remains to measure the distances to these objects in nearby hosts as well as in galaxies where  SNeIa are also observed. 

Three main anchors (nearby hosts) are employed in the state-of-the-art \citep{shoes22} cosmic distance ladder calibration. These are%
\footnote{The additional 145 Cepheids in the Small Magellanic Cloud (SMC) and 55 Cepheids in M31 are only used for deriving the metallicity correction term of the period-luminosity relation. In the SMC the challenge is the lack of Hubble Space Telescope (HST) observations of the Cepheids (while the differential distance between SMC and LMC is well known, see \cite{Graczyk2020,Breuval2021}). In M31 the problem is the lack of independent geometric anchoring methods to determine its distance to a high enough precision.}
\begin{itemize}
    \item 83 Cepheids in the Milky Way \citep{Riess2018a,Riess2021MW}, to which the distance has been measured using parallaxes\footnote{The parallax measurement is based on the angular displacement of a nearby object caused by the shift in observing position from Earth as Earth travels along its orbit around the Sun.} from HST WFC3 (8 Cepheids) and Gaia EDR3 (75 Cepheids) at around $\sim 1\%$ precision.
    \item 669 Cepheids in NGC4258 \citep{Yuan2022,Reid2019}, to which the distance is well determined from the water masers orbiting its central black hole. These masers emit light of a well-known frequency whose Doppler shift can lead to precise determination of their movement. By combining masers far from the center to determine the velocity and angular extent of the accretion disk and masers close to the center to determine the velocity drift (and thus centripetal acceleration), the distance to the focal point of the Keplerian orbit can be determined. Using Very Long Baseline Interferometry (VLBI) then gives a 1.3\% precision of the distance \citep{Reid2019}. Note that this is a distance measured geometrically,  to a cosmological extent of around $\sim 7.6$ Mpc. Further discussion of masers as distance indicators can be found in \cref{sssec:single_rung}.
    \item 70 Cepheids in the Large Magellanic Cloud (LMC), the distance to which has been determined using detached eclipsed binaries \citep{Pietrzynski2019}. In these binary systems, the suppression of flux when an eclipse is occurring allows for the determination of the absolute sizes of the objects. Combining this absolute size with the angular size as determined from the observed flux and calibrated surface brightness\footnote{As opposed to the intrinsic luminosity (which combined with the observed flux would directly yield the luminosity distance), the surface brightness is a luminosity per angular area, and thus only allows for a measurement of the angular extent of an object when combined with the observed flux.
    The surface brightness vs temperature relation was calibrated geometrically for Red Giants using measured interferometry of local Red Giants, a key step to enable increased precision  and accuracy in distances.
    } (using the surface brightness color relation) allows for the determination at the 1.5\% level of their overall distance.
\end{itemize}

 \paragraph{Second rung of the ladder} The three independent anchors mentioned above calibrate the Cepheids' period-luminosity relation zero-point and slope (and metallicity correction), which then allow for observations of Cepheid periods in the distant hosts of supernovae (and other objects, see below) to be converted into distance measurements. These, in turn, are then used to calibrate the next step of the distance ladder, which involves SNeIa. 
Measuring the Cepheid fluxes in the previous rung and this one using the same telescope, instruments, filters and zero-points is key to negate systematic errors (more on this below).

SNeIa are caused by binary systems with an accreting carbon-rich white dwarf, whose mass passes the Chandrasekhar threshold at which Carbon fusion suddenly becomes possible, and the ensuing sudden release of energy causes the supernova explosion; a binary white dwarf merging where the sum is near the Chandrasekhar mass is another possible scenario.  Since the explosion is caused by objects of roughly the same mass with the same underlying mechanism, the observed light curves are remarkably similar. Once calibrated, the shape of this light curve can then be used to reconstruct the influence of the astrophysical environment (such as metallicity) as well as the intrinsic standardized luminosity. This is typically done using light curve models such as SALT 2/SALT 3 \citep{Kenworthy2021}. 

\paragraph {Third rung of the ladder} SNeIa are abundant enough that, once their  relation between the intrinsic luminosities and measured light curves is calibrated and their distances can be inferred, they  can be used in the third rung of the ladder to measure the distance-redshift relation very accurately in the Hubble flow.

The main uncertainty that propagates into an $H_0$ measurement is the calibration  of the distances to SNeIa supernovae  hosts, and is limited by the relatively small number of supernovae hosts in the range of Cepheids\footnote{Recall that  this corresponds to distances below about 40 Mpc where the overall rate of SNeIa is about one per year. The calibration error scales roughly as $6\%/\sqrt{N}$ with $N$ being the number of supernovae hosts with known distances.} (and thus with known distances). 
For the state-of-the-art  measurement of \cite{Riess2022}, the number of SNeIa hosts with Cepheids distances has been increased to 42 (Pantheon+ analysis, \cite{Solnic2022,Brout2022}). Given the supernovae rate in the relevant volume this number is unlikely to increase significantly in the near future. 

In the third rung of the ladder, the supernovae beyond these hosts can then be used to determine luminosity distance far out in the Hubble flow via 

\begin{equation}
    D_L(z) = \frac{1}{H_0}(1+z) \int_0^z \frac{dz'}{E(z')}
\end{equation}
which allows one to jointly determine the Hubble parameter $H_0$ and the unnormalized expansion rate $E(z) = H(z)/H_0$ -- the latter is usually compressed under the assumption of a $\Lambda$CDM model into the matter fraction $\Omega_m$\,. 
For a smaller redshift range up to $z<0.15$ the distance redshift relation is expected to be approximately linear, hence a cosmographic expansion
\begin{equation}
    \log D_L(z) \approx \log z \left[1+\frac{1}{2}(1-q_0) z -\frac{1}{6}(1-q_0-3q_0^2+j_0)z^2\right] - \log H_0
    \label{eq:cosmographic}
\end{equation}
with $q_0 = -0.55$ and $j_0=1$ is used in \cite{Riess2022} to find a value of the Hubble constant of $H_0 = 73.04\pm 1.0 \mathrm{(stat)}\pm 0.3 \mathrm{(sys)}\,\hunit$.
\footnote{The value of $j_0=1$ is predicted from the $\Lambda$CDM model, with $q_0\approx-0.55$ being a good approximation for $\Omega_m \sim 0.3$.  Analyses with varying   $q_0$ determined only from higher redshift SNeIa (marginalizing over cosmographic fits) are also performed in \cite{Riess2022}, and the result is almost unchanged.} It remains to be stressed that there is a small dependence on the form of \cref{eq:cosmographic} in this last step. While typically $E(z)$ is a relatively flat function at $z < 2.5$ and can thus be well approximated by a cosmographic expansion at low redshifts, there could in principle be models with a rapid late time transition of dark energy, in which the translation of luminosity distances to the Hubble parameter should be performed taking the special expansion rate $E(z)$ of these models into account.

One important aspect of the determination  of $H_0$ from the distance ladder is the correction for peculiar velocities%
\begin{marginnote}[]
\entry{Peculiar velocities}{Deviations of recession velocities (and thus redshifts) from the pure Hubble flow. They are sourced by the gravitational pull of matter inhomogeneities.}
\end{marginnote}%
which are deviations from the pure Hubble flow. In our inhomogeneous Universe the measured redshift of a cosmological object is not purely due to the expansion. It also includes a component caused by the Doppler effect from the peculiar velocity, which is sourced by the gravitational pull of inhomogeneities (such as galaxies falling towards the centers of cosmic filaments, or incoherent velocity dispersions in high-density regions). 
While the peculiar velocities of objects are typically around the range of $\mathcal{O}(10^2\mathrm{km/s})$, the Hubble flow recession velocities grow with the distance. As such, the relative importance of the peculiar velocity component compared to the Hubble flow decreases with distance (and this is the reason why a minimal redshift cut of $z \gtrsim 0.02$ is used to reduce sensitivity to peculiar velocities). Separating the components of redshift due to expansion and peculiar motion in the nearby universe ($z < 0.1$) is critical to provide an unbiased measure of the Hubble constant. Multiple methods are  employed to correct for peculiar velocities (see for example \cite{Peterson_2022} for a recent review). With the treatment of peculiar velocities that most reduces the uncertainty these affect $H_0$ only at the $0.1\,\hunit$ level, while the overall correction in principle affects $H_0$ at the $0.5\,\hunit$ level (between correcting for peculiar velocities and neglecting them completely).

Recently the use of machine learning has allowed for additional parameterization of SNIa spectroscopic information (using the deepSIP framework of \cite{Stahl2020}), which has been used in \cite{Murakami2023} to decrease uncertainties on the Hubble parameter by~14\%.
This leads to the most precise Hubble constant determination from the local distance ladder to date of
\begin{equation}\label{eq:H0cepheid_boosted}
   [1.a]\,\,\, H_0 = 73.29 \pm 0.85 \mathrm{(stat)} \pm 0.3 \mathrm{(sys)} \,\hunit
\end{equation}
which is at a remarkable $5.7\sigma$ level of tension with the value determined from Planck (see \cref{ssec:CMBmain}).

When observing the same distance ladder in the near-infra-red (NIR), for 19 SNIa hosts and 57 SNeIa in the Hubble flow, \cite{Galbany2022} obtain a relatively precise determination of $H_0 = 72.3\pm 1.4 ({\rm stat)} \pm 1.4 ({\rm sys)}\,\hunit$ (with only small variations for the H or J observation band). Furthermore, recent James Webb Space Telescope (JWST) observations of extra-galactic Cepheids \citep{Riess2023} -- with photometry measured independently from the SH0ES collaboration -- confirm the findings from the Hubble Space Telescope. This implies that systematic errors in HST Cepheid photometry (especially crowding\begin{marginnote}[]
\entry{Crowding}{When multiple unresolved background objects 
confound the measurement of a foreground object.}
\end{marginnote}%
from other stars) are bounded and thus cannot play a significant role in  shifting the Cepheid-based $H_0$ determination.

\subsubsection{A two-rung Cepheid distance ladder}\label{ssec:two_rung}
One of the possible criticisms of the standard distance ladder approach is the reliance on indirect calibration (of supernovae on  Cepheids in far away hosts). However,  
compelling evidence is emerging that this indirect calibration is not at the heart of the Hubble tension. In particular, in \cite{Kenworthy2022} the Hubble parameter is obtained  by completely removing the supernovae from the ladder, and only using a two-rung distance ladder based on the anchors and the Cepheids.

The result is a striking argument against problems in supernovae or the indirect calibration step causing the tension. While giving up on the supernovae necessarily reduces the lever arm and increases the uncertainty of the measurement, 
the result of $H_0 = 73.1^{+2.6}_{-2.3}\,\hunit$ is still precise enough to give a 2.6$\sigma$ tension with the Planck result. The systematic error budget is dominated by the uncertainty from the estimation of peculiar velocities, which is to be expected, as the Cepheids alone do not reach as far into the Hubble flow and are thus more strongly impacted by peculiar velocities.

\subsection{Alternative distance ladders}\label{ssec:alternative_dist}

\subsubsection{Tip of the Red Giant Branch}
\label{sec:TRGB}
The standard distance ladder based on Cepheids and supernovae presented above yields  one of the most precise determinations of the Hubble constant. 
However, because of the profound implications of the Hubble tension, it is of utmost importance to provide confirmation or corroboration by independent approaches relying on different observables and different  physical processes. 

The most well-known alternative distance ladder is based on measurements of another critical phenomenon of stellar objects. When red giant stars continue burning hydrogen to helium, their degenerate helium core slowly grows until at some point the pressure is high enough for helium fusion to begin (helium flash). Since that pressure is mostly a function of the core mass, which has to pass a critical threshold (about 0.5 solar masses), the overall luminosity and color of stars close to this point are remarkably similar.  This results in a sharp cut in the population of red giant stars above a given intrinsic luminosity in the  Hertzsprung Russel diagram. This discontinuity in the luminosity function is called the tip of the red giant branch (TRGB).

It has recently been shown that the TRGB can be used as a standard candle reliably enough to calibrate the cosmic distance ladder \citep{TRGB19}. 
The TRGB has the advantage of being measurable in   all galaxy types and thus many more galaxies~$\mathcal{O}(500)$ compared to Cepheids~$\mathcal{O}(100)$. However, the TRGB luminosity is about 10 times fainter than long-period Cepheids and thus can yield reliable  distances only out to $\sim$ 20 Mpc. 
Unlike Cepheids which are individual objects, it is not possible to  directly measure individual parallaxes of TRGB in the Milky Way, thus calibration (or standardization) of the TRGB needs to be done in some other way.

Moreover,  TRGB detection  requires fitting for a more or less sharp cut in the luminosity function, which is further muddled by asymptotic giant branch stars which can be rather close in magnitude. This is typically performed using smoothing, Sobel filters, Kernel density estimators, or by fitting a broken power law, though other methods have also been put forth.

The TRGB-based distance ladder first produced competitive results for the Hubble constant in 2019 following the effort of the Carnegie Chicago  Hubble project CCHP \citep{TRGB19}. It is completely independent of Cepheid anchors,  but it still relies on SNeIa for the second rung of the distance ladder to provide an $H_0$ measurement.   At the moment of writing two independent groups have produced state-of-the-art determinations of $H_0$ using this approach.
One (the original) group  (CCHP, \cite{TRGB20}) obtains $H_0 =69.8 \pm 0.8 {\rm (stat)}\pm 1.7{\rm (sys})\,\hunit$  [1.$b$], while the other group  (EDD, \cite{anand22}) obtains a higher value of 
$H_0=71.5\pm 1.8\,\hunit$.

The main TRGB-related differences between the analyses of the two groups are the determination of the TRGB magnitudes in the anchors (in particular for NGC4258) and the precise method used to determine the cut in the luminosity function. However, the CATS collaboration \citep{CATSScolnic23} suggests that a difference of up to $\sim 2 \,\hunit$ might also stem from the precise supernovae sample and its correction for the peculiar motion of the host galaxies, with the specifics of the TRGB calibration method accounting for a shift of only $\sim$ 1.4 $\,\hunit$.
The CATS collaboration reanalysis of TRGB-based distance ladder \citep{CATSScolnic23} 
obtains $H_0=73.22\pm 2.06\,\hunit$, using an unsupervised learning algorithm (minimizing field-to-field dispersion, described in \cite{Wu2023,Li2023}) to find a relation between the tip contrast (ratio of stars above/below the tip) and the TRGB absolute magnitude. 

As such, one of the fundamental issues to be settled  
for the use of the TRGB as a consistent standard candle is the precise determination and definition of the tip.
Out of the 19 supernovae hosts with known TRGB distances the Cepheid-based distance is known for 7 of them. \cite{Riess2022} demonstrates that  using the supernovae hosts in common between the TRGB and the Cepheids and calibrating only on the common anchor NGC4258, a very consistent calibration with $0.00\pm 0.03$mag for both the analyses of \cite{TRGB20} and \cite{anand22} is found.

At the moment we should probably conclude that due to the TRGB-based distance ladder being a somewhat newer approach to $H_0$, the dust has not entirely settled. However, the importance of having a distance ladder and an $H_0$ determination independent of Cepheids cannot be overstated. A red teaming approach\begin{marginnote}[]
\entry{Red teaming}{The practice of viewing a problem from an adversary or competitor's perspective (a.k.a. playing the Devil's advocate).}
\end{marginnote}%
to both routes to $H_0$ may be a possible way forward. 
In the meantime approaches such as \cite{Uddin23} might help bracketing the extent of possible \enquote{unknown unknowns}.

\subsubsection{Tip of the Asymptotic Red Giant Branch}
Beyond the TRGB, other standard candles are being investigated such as the tip of asymptotic red giants \citep{Madore2020,Lee2021}. These objects undergo a similar physical mechanism as that generating the TRGB, except that the core is one made up of carbon, and the sudden change in the star's brightness appears when sudden carbon burning starts in the core. These are typically measured in the near-infrared in the J band and provide competitive accuracy \citep{Lee2022}. There is also a variation on this theme, using Carbon stars \citep{Parada2023,Madore2023}, which are stars close to this tip of the asymptotic red giant branch, but experience carbon from the core being dredged up towards the surface, which gives them a signature redder taint (shifting them into the infrared). Since these stars are redder than the whole of the asymptotic giant branch, there are not a lot of other star types crowding the measurement, and their distribution is closer to a peak rather than a cutoff, yielding a  clearer signature. The use of the tip of the asymptotic red giant branch or the Carbons stars has not yet led to competitive constraints on $H_0$ due to the lack of observations in supernovae hosts, but with the recent launch of the James Webb Space Telescope (JWST) this is expected to change soon, potentially delivering another local calibration of the Hubble constant.

\subsubsection{Other variable stars}
In principle, it should be possible to link intrinsic luminosity to the period of  any type of variable star with a well-known mechanism of pulsation. However, in practice, the use of these pulsating standard candles is not trivial. Apart from Cepheids, for the second rung of the distance ladder, only Mira variable stars have led to a competitive measurement of the Hubble constant at this point (see below). While RR-Lyrae stars and type two Cepheids also show standardizable period-luminosity relations, the systematic effects plaguing these two pulsating objects have so far represented  a major  roadblock. While the former suffer from strong metallicity dependence\footnote{Despite this, \cite{2013MNRAS.435.3206D} has released a measurement in 2013, reaching $H_0 = (80.0\pm3.4)\,\hunit$, though this result is simply an up-scaling based on a different calibrator distance instead of a full distance-ladder measurement based on RR-Lyrae stars. It should also be mentioned that there has been a lot of recent progress in this field, for example in 
\cite{Mullen2021,Garofalo2022,Mullen2022,Mullen2023}.} (and a corresponding lack of accurate parallax measurements), the brighter type two Cepheids suffer from comparatively low abundance (due to faster evolutionary timescale), which makes detailed investigations of their pulsation properties harder \citep{Bhardwaj2020}.

Instead, Mira variables (which have a tight period-luminosity relation observable only in the near-infrared) are more regular and allow for another Cepheid-independent distance ladder using essentially the same principles as employed for the Cepheids. Compared to Cepheids, the Mira variables have lower-mass progenitors and are more available in different galaxy types and further out in the halo. However, critically, they have been found only in a single supernova host (NGC1559, SN Ia 2005df) so far. Based on this single host, \cite{Huang2020} find
\begin{equation}
    [1.c]\,\,\, H_0 = 73.3 \pm 4.0 \,\hunit
\end{equation}
showing that the Hubble constant is measurable, though more supernova hosts and anchored Miras need to be identified. Anchors currently include the LMC, NGC4258, and recently the Milky way by \cite{Sanders2023} who finds $73.7 \pm 4.4\,\hunit$.

\subsubsection{Type II supernovae}\label{ssec:snii}
To strengthen the ladder, apart from changing the calibrator object, it might also be interesting to change the final faraway objects in the Hubble flow: after all a possible (unknown)  systematic in these faraway objects could in principle bias the Hubble determination. One promising avenue is the use of core-collapse supernovae instead of SNeIa.

Core collapse supernovae occur when a massive star fuses the last of its material that can undergo fusion with positive energy gain. Typically the star is layered with heavier elements such as Silicon, Magnesium, and Neon fusing closer to the core, and lighter elements such as Oxygen, Nitrogen, and Carbon fusing further out, surrounded by a Helium and Hydrogen shell. The innermost fusions produce Iron and Nickel, which do not produce further energy upon fusion. As such, when the fusion products around the core are mostly burnt up and stop providing sufficient gravitational support against a collapse, the whole star implodes. The intense gravitational pressure condenses the inner core beyond the electron degeneracy limit, creating a large amount of neutrinos in the proton-electron fusion. Only when the neutron degeneracy pressure limit is reached is the collapse halted, and reflected.\footnote{For stars above the Tolman Oppenheimer Volkoff limit, there is the possibility that even neutron degeneracy pressure is not enough, and instead of a neutron star remnant a black hole is formed instead. In this regime it appears possible but not necessary to generate a supernova (see \cite{Burrows2021,Gullin2022}), though the research in this direction is ongoing.} 
The reflected shock is accelerated by the outflowing neutrinos and results in a massive explosion carrying away enough energy to be visible from cosmological distances. Depending on the amount of hydrogen and helium in the outer shell, the supernova is classified as type Ib/Ic or type II.

In \cite{deJaeger2022} the authors describe a distance ladder based on Cepheids and the TRGB using supernovae of type II instead of the commonly used type Ia, based on \cite{deJaeger2020a,deJaeger2020b}. These are typically fainter but more abundant than the type Ia supernovae, and, most importantly,  must be standardized (for example through their expansion velocity and color). Using 13 calibrator galaxies with either known TRGB or Cepheid distances, and 89 supernovae in the Hubble flow, they find a combined measurement of $H_0 = 75.4^{+3.8}_{-3.7}\,\hunit$ (with an approximate $\pm 2\,\hunit$ systematic uncertainty), with only a slight ($<1\sigma$) difference depending on whether they use exclusively Cepheids or the TRGB calibrations.
While error-bars are relatively large ($\sim 5$\%), this finding corroborates that of \cref{ssec:two_rung}: there is no strong evidence that SNeIa are the source of the $H_0$ tension. 

\subsubsection{Megamasers -- A single-rung distance ladder}\label{sssec:single_rung}
A single rung distance ladder would completely avoid the calibration of different distance indicators and thus negate any systematic errors arising from the calibration process. 
This has recently become possible as, 
in the last decade, masers have been detected even to cosmological distances of hundreds of megaparsec. This approach is limited by the small number of rare edge-on disks suitable for distance measurements and by the fact that the maser systems at $\sim$ 10-100Mpc  distances  are faint for the sensitivity of the radio facilities available. 

In the context of the Megamaser Cosmology Project, \cite{Pesce2020} uses 6 maser systems up to $\sim 130$Mpc (well into the Hubble flow) to determine the Hubble constant from their distances and recession velocities. Thus $H_0$ is obtained in a single step, requiring no external calibration.

The final result is (as expected) dependent on the precise assumption of the peculiar velocities of these objects. Assuming zero means of the peculiar velocities (i.e., no peculiar velocity corrections) with a 250km/s uncertainty, the overall result is $H_0 = 73.9 \pm 3.0\,\hunit$ (with a similar value obtained when the peculiar velocity averaged over the associated galaxy group is used). If instead modern galaxy flow models 
are adopted, the authors find slightly lower (but still consistent) values for the Hubble constant, such as $71.8^{+2.7}_{-2.6}\,\hunit$ for the CosmicFlows model from \cite{Graziani2019}.
It is important to note that the megamaser $H_0$ measurement is independent of the  distance ladder. The converse is not necessarily  true as some $H_0$ measurements use NGC4258 as one of the anchors. But \cite[Tab.~2]{Pesce2020} excluding NGC4258 obtains  (with a galaxy flow model)
$H_0=72.1\pm  2.7\hunit$ [3.$a$]. This measurement is  independent of the distance ladder as no data is in common in either direction (as such it could belong to \cref{sec:noladder} below).

\subsection{Beyond astrophysical objects}
Building distance ladders based on individual astrophysical objects is not the only way to infer the Hubble constant. Instead, even galactic properties can be used for standardization. We treat surface brightnes fluctuations in the main text, other approaches can be found in supplementary material of  \cref{ssec:otherladders}.

\subsubsection{Surface brightness fluctuations}\label{ssec:sbf}
The idea of the surface brightness fluctuation (SBF) measurement is based on the empirical calibration\footnote{A theoretical calibration based on star formation models has also been used in the past, but does not achieve the same systematic uncertainties as the empirical calibration.} of a tight relation between the intrinsic magnitude of the observed SBF and the color of the object. This allows the color measurements of the host to determine the intrinsic magnitude, while the SBF measurements are used to determine the observed magnitude. Both of these combine to give a distance estimate, which can be combined with the redshift measurement of the object to directly measure the Hubble parameter (as in \cite{Blakeslee2021}) or to calibrate the supernovae absolute magnitude in the given objects (as in \cite{Khetan2021}). The zero-point of the color-magnitude relation currently has to be calibrated using external information, such as using Cepheid or TRGB distances.

The SBF measurement is challenging, with a review of the steps outlined in \cite{Cantiello2018,Moresco2022,Cantiello2023}. While the mean surface brightness (observed flux per \emph{angular} area) is roughly independent of distance, the Poissonian fluctuations thereof (caused by the finite numbers of stars) are not. A factor of $2$ in distance will result in only $1/4$ of the observed flux for a given physical area, but a given angular area will also correspond to $4$ times the physical area, making the surface brightness approximately distance independent. However, the standard deviation of the Poissonian fluctuations of the surface brightness will only increase by a factor of $\sqrt{4}=2$ from an increase of factor $4$ in the physical area. The ratio of the SBF variance and the mean surface brightness then constitutes a distance-dependent SBF flux (which is simply the flux-weighted average intrinsic flux of the objects).  
To actually obtain the SBF one needs to subtract the main galaxy mean flux contribution, mask any unmodeled residuals and interloping objects to obtain a clean noise map, model the impact of background galaxies and the globular cluster luminosity function to obtain the residual noise power spectrum, which can be decomposed finally into a flat white-noise component (caused for example by the detector) and a non-flat component smeared by the point spread function (PSF) of the measurement. Subtracting other PSF-dependent noise also arising from residual globular clusters and background galaxies, one finally obtains the SBF contribution. This contribution can then be converted to the observed magnitude using a simple linear relation that has to be calibrated in local galaxies. This procedure requires multiple stages of calibration, which is why the measurements are currently dominated by systematic rather than statistical uncertainties.

In \cite{Khetan2021} the authors use the SBF distances to calibrate 24 SNeIa hosts and only 96 SNIa in the Hubble flow ($z<0.075$), finding $H_0 = 70.5 \pm 2.4 {\rm (stat.)} \pm 3.4 {\rm (sys.)}\,\hunit$ (with the SBF zero-point calibrated on nearby Cepheids, effectively resulting in a four-rung distance ladder). However, from an improved measurement of the LMC distance from \cite{Pietrzynski2019}, the mean value of this analysis would rise to $71.2\,\hunit$ \citep{Moresco2022}. This measurement is superseded by \cite{Garnavich2023}, who measure $H_0 = 74.6\pm0.9\pm2.7\,\hunit$ [1.$d$], using infrared measurements of the SBF in 27 SNIa hosts and 816 SNIa from the Pantheon+ sample ($z<0.25$). This appears to independently confirm a preference for a relatively high $H_0$ value even though the SNIa and their host galaxies are very different (differing in lightcurve duration and host mass) and between SBF-calibrated and Cepheid-calibrated hosts. It should be noted that adopting SNeIa only in these fast-declining massive galaxies and refitting the SALT2 SNIa standardization parameters, \cite{Garnavich2023} find $H_0 = 73.3 \pm 1.0\,\hunit$ (without systematic errors), wich central value agrees well with the Cepheid-based calibrations.

Instead, in \cite{Blakeslee2021} the SBF measured in the infrared of 63 galaxies up to distances of $\sim 100$Mpc (with intrinsic magnitude zero-point calibrated on the TRGB/Cepheids) is then directly used to infer the Hubble constant (three-rung distance ladder using SBF instead of SNIa). This results in a measurement of $73.3 \pm 0.7 {\rm (stat.)} \pm 2.4 {\rm (sys.)}\,\hunit$ (with consistent results between calibration on TRGB or Cepheids), with the result not strongly dependent on the assumed model for the peculiar velocities (using similar corrections as in \cref{sssec:single_rung}).

The future of this probe is rather bright, with JWST expected to deliver a wealth of IR SBF measurements \citep{Cantiello2023}.

\subsection{Putting it all together: some thoughts on systematics and risk assessment}

\begin{figure}
    \centering
    \includegraphics[width=\textwidth, height=8cm]{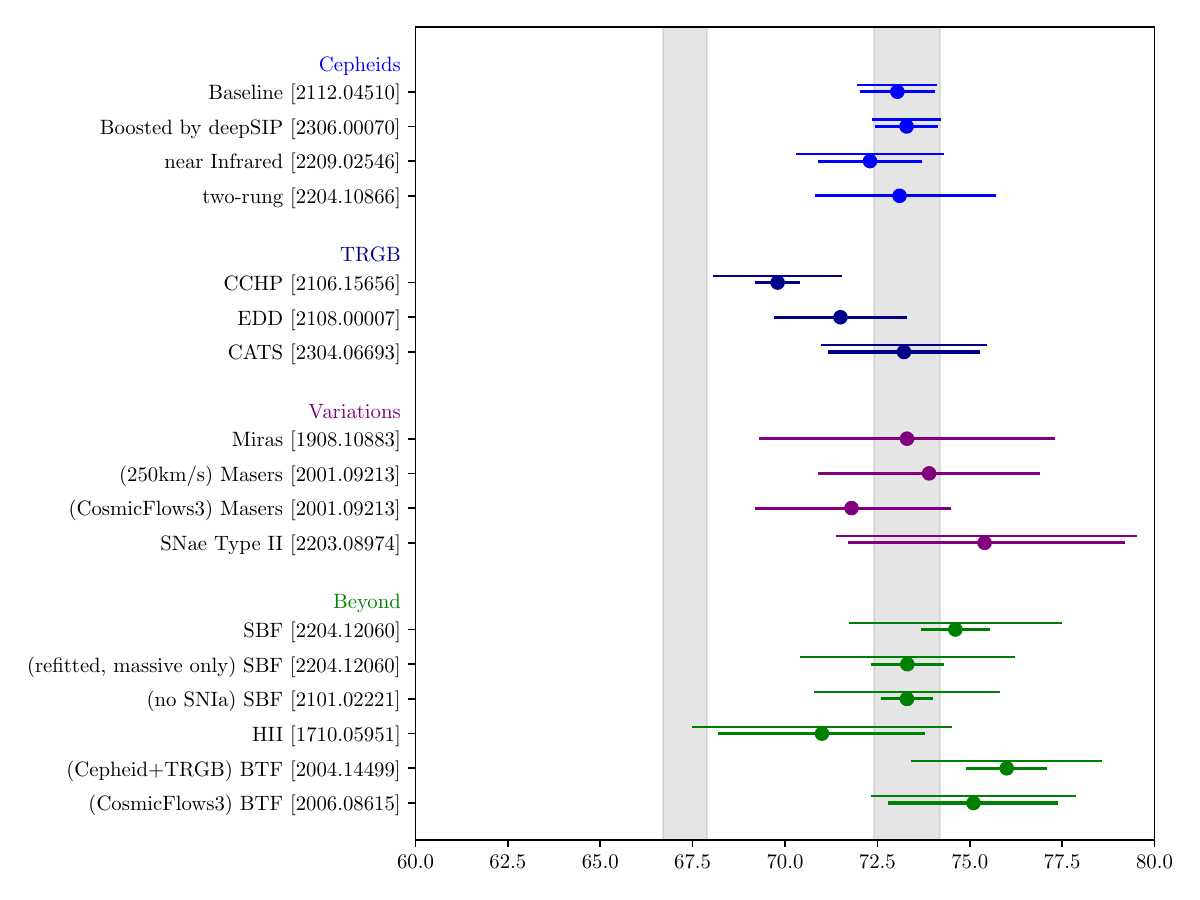}
    \caption{Comparison of  the  direct distance ladder determinations of $H_0$ discussed in \cref{sec:distance} (all error bars are 1$\sigma$, 68\% CL). If a given analysis includes a determination of the systematic uncertainties, they are shown as a second (vertically offset) error bar for which the systematic and statistical errors have been added in quadrature. Hereafter the two shaded areas denote the two \enquote{camps} i.e.,\cite{PlanckParams2020} and \cite{Riess2022}.}
    \label{fig:local}
\end{figure}

The family of approaches to measure $H_0$ based on a distance ladder measure directly the distance-redshift relation. These,  by construction, are limited to  relatively low-redshift objects yielding a \enquote{local} $H_0$ measurement. While each singular determination relies on several assumptions and is potentially affected by unknown systematic errors, the consistency of the multitude of these different methodologies signals that these are mostly under control. Finding and mitigating  
systematics and  developing new, independent,  avenues to the Hubble parameter is crucial: the last decade has seen a significant (possibly heroic) effort in this direction. However, great care should be exercised when taking the consistency of the local determinations of the Hubble constant as an argument, as many of the approaches are not as independent as they might superficially appear.

One possible approach was proposed by \cite{baccus}: to find a consensus $H_0$ constraint,  all different measurements, even those that are in tension,  could be combined (in a somewhat agnostic way) while the resultant  uncertainty  adjusts to reflect the spread among determinations.    

Another approach is to  try to recognize the synergies and complementarities among the different methodologies.  In our view each of these local $H_0$ measurements can in principle  be further sub-divided into independent intermediate results (which we refer to as \enquote{modules}). In many cases these \enquote{modules}  can be compared and their  careful comparison can be used to put a ceiling to possible systematics (known unknowns or even unknown unknowns). This is in spirit similar to the red teaming work mentioned in \cref{sec:TRGB}. Below we sketch how such an approach may work. 

\subsubsection{Anchors}
One possibly underappreciated aspect is that of the anchors. Currently the various distance ladders of \cref{sec:distance} are mostly built on three main anchors: the Milky Way parallaxes, the eclipsing binaries of the Large Magellanic Cloud, and the maser distance to NGC4258. The latter is often used to anchor both the TRGB and the Cepheid-based distance ladders, which are currently the ones yielding the lowest uncertainties on the Hubble constant. Since these objects anchor the entire measurement, any issue with the distance to these objects would directly result in a systematic shift of the Hubble parameter. At this point, there is no indication of a systematic shift of all three anchors (see \cite{Efstathiou2020,Efstathiou2021} for a discussion on this possibility), and the constraints from the individual anchors are  all mutually consistent \citep{Riess2022}. Still, given the pivotal role of these anchors in the overall measurement, and the low number of anchoring objects, developing independent approaches to set the overall scale of the measurement and  to further bracket any possible systematic shift remains an important priority. 

\subsubsection{Cepheids} Cepheids are, at the moment, the preferred distance indicators for the cosmic distance ladder as they are extraordinarily bright and thus can be seen at large distances $D>10$ Mpc. Moreover, the distance per star is intrinsically precise  ($\sim3$\% per star). Of course, this assumes that Cepheids are good distance indicators  and that therefore  their key properties do not change  along the distance ladder. It should be stressed that this is  not only motivated physically from  our understanding of stellar theory, but, in practice, it is confirmed empirically.

In particular, the fundamental assumption underlying this determination of the Hubble constant rests on the cosmological principle: Cepheids in our local environment within about $10$Mpc obey the same period-luminosity relation as high-redshift Cepheids beyond about $10$Mpc (for finding enough supernovae hosts with Cepheids distances).  There is no indication of any such  \enquote{break} in the ladder. 

In fact, the number of suitable  hosts that harbour both Cepheids and supernovae  with the \enquote{right} properties  (the second rung of the ladder) is the major limitation  to the precision on $H_0$ in this step ($\sigma_{H_0}\sim 6\%/\sqrt{N}$ with $N$ the number of such hosts). This number has been  steadily  increasing from $N=19$ in 2016 to  $N=42$ in 2020, significantly reducing this uncertainty. 
To \enquote{break}  this step along the ladder would require postulating that Cepheids  calibrators at $D<10$Mpc are intrinsically different from Cepheids at $10<D/{\rm Mpc}<40$ even if they  have the same luminosity, mass, and pulsation period within a small interval.

It should be noted that this is a well-established area of stellar evolution and there is no strong evidence for any deviations from predictions. One particular example to validate our understanding of the physical mechanism can be seen in \cite{Riess2022} where the observation of the \enquote{Hertzsprung Progression} and the detection of resonance between pulsating modes are shown, demonstrating the universality of the physics producing pulsating standard candles, and their consistency along the distance ladder.
Moreover, the Cepheids used in the ladder are quite uniform in terms of luminosity and mass. There is evidence for some metallicity dependence, which is expected and can be calibrated (see for example \cite{Romaniello2022,Breuval2022} and references therein), although a misunderstanding of the  dependence of the period-luminosity relation on the stellar environment could in principle be problematic.

Cepheids have now been consistently measured in independent instruments and frequency bands, and the standard distance ladder has been consistently constructed using the same photometric instrument, hence issues with zero-point errors seem relatively unlikely. The possible extent of systematics with parallaxes determinations to nearby Cepheids has been increasingly reduced, starting with HST spatial scanning \citep{Riess2018}, then including Gaia parallaxes \citep{Riess2021MW,Riess2022b}. 

Finally, the Cepheids period-luminosity relation and thus distance calibration analysis has been reproduced -- fully independently from the SH0ES team -- for host galaxy NGC5584, starting from raw data \citep{Javanmardi2021}. They confirm SH0ES findings  while  employing different tools for  the  photometric analysis and a completely different method for  light curves.

The major problem for HST Cepheid photometry is believed to be crowding, which is confusion with other stars. Long period, bright Cepheids  that can be seen at large distances $D>10$Mpc are particularly affected by this problem.  Recent re-observations of distant Cepheids with JWST \citep{Yuan2022b,Riess2023} indicate that any possible effect of crowding on the $H_0$ determination is negligible at best and that HST Cepheid photometry can't be \enquote{biased bright} enough to mitigate the Hubble tension. The dust extinction could in principle be another worry for the Cepheid calibration, but consistent measurements have been carried out in the near-infrared and mid-infrared (using JWST), where the dust extinction is much lower, showing no significant deviation.

For the standard distance ladder, the sum uncertainties from the three anchors and the uncertainties for the small number of supernova Ia hosts (19 before 2020, now 42) are dominant contributions (85\%) to the overall uncertainty.

\subsubsection{TRGB}
The TRGB determination is particularly impacted by the precise determination and definition of the \enquote{tip}, which does have a significant impact on the inferred Hubble constant value, as demonstrated in \cite{CATSScolnic23}. The natural definition through the fastest change in luminosity function adopted by \cite{TRGB20} gives different results from an approach that attempts to maximize the contrast between numbers of objects above and below the tip \citep{CATSScolnic23}.

\subsubsection{Megamasers}
It is interesting to explicitly bring attention to \cref{sssec:single_rung} on the Megamaser cosmology project, which is a direct single-rung determination of the Hubble parameter based on maser distances. While the measurement has large systematic errors stemming from the assumptions about the exact peculiar velocities of the underlying objects, it shows that completely independent determinations are possible. These end up lying in a somewhat high-$H_0$\, region.

\subsubsection{Supernovae}
In the last rung of the ladder supernovae are the preferred standard candles which are well in the Hubble flow. The  distance-redshift relation as obtained from Supernovae is in principle affected by cosmological bulk flows at low redshifts and cosmology-dependence at higher redshifts. The applied  redshift cuts reduce the dependence on both down to a level often much smaller than statistical uncertainties. When approaches of modeling the peculiar velocity field such as \cite{Tully2023} are applied, they  are typically based on a large compilation of different methods, which makes propagation of systematic errors challenging but, as discussed above, the overall correction can be kept small.  

\subsubsection{Alternative ladders} It should be noted that most of the alternative  distance-ladder determinations of the Hubble constant are calibrated using the Cepheids or the TRGB as calibrators, thus sharing the same anchor and calibrator systematics. This is true for the supernovae of type II of \cref{ssec:snii}, and all the alternative ladders from \cref{ssec:alternative_dist}. It should, however, be highlighted, that the consistency of these alternative determinations using the same calibrator makes it rather unlikely that supernovae systematics are at the heart of the Hubble tension, as also evidenced in the two-rung distance ladder of \cref{ssec:two_rung}. None of these alternative distance ladders alone currently has sufficient precision to give a large Hubble tension $>3\sigma$, but their remarkable consistency shows that possible systematics are more likely to be either at the level of the calibrators (Cepheids or TRGB) or the anchors (see above).

\subsubsection{Summary} A shift by $\sim 3-5\,\hunit$ in the locally determined value of the Hubble constant (required to solve the tension) would require several systematic effects to have been all underestimated  and to all combine in the same direction, which seems increasingly more unlikely.

It seems clear that the control of systematic errors is now crucial. Unfortunately, different authors/teams seem to treat systematic errors differently. In some cases combined systematic errors are propagated in the final result separately from the statistical errors, sometimes their breakdown by source is reported, while in other cases systematic and statistical errors are combined into a final error-bar (and some cases even lack a systematic uncertainty altogether). Given the exquisite precision and accuracy of the measurements, going forward a transparent breakdown and traceable treatment of systematic uncertainties would greatly facilitate progress.

\section{INVERSE DISTANCE LADDER(S)}\label{sec:invdist}
\Cref{sec:distance} focused on determining the Hubble constant locally through astrophysical processes. However, being a global parameter of the cosmological model, $H_0$ can also be inferred through its indirect impact on early-time processes. 
Early-time here refers to the times before recombination ($z \gtrsim 1100$), which corresponds to approximately the first 400\,000 years of the Universe.
These early-time processes are typically sensitive to the physical energy densities of the various species present in the Universe, which,  through the Friedmann equations, relate to the absolute Hubble parameter. 
If the relative expansion rate between the early times and the present-day Universe  can be simultaneously determined (or is fixed by the cosmological model), then  the Hubble parameter can be inferred from  observations of these early-time processes. The most notable of these processes is the generation of the cosmic microwave background (CMB), the baryonic acoustic oscillations (BAO)%
\begin{marginnote}[]\entry{BAO}{Baryon Acoustic Oscillations}\end{marginnote}%
of the early plasma, which is imprinted both in the CMB and the late-time clustering of large scale structure, and the light element abundances of big bang nucleosynthesis (BBN).
A large part of the constraining power for $H_0$ that these processes offer is rooted in our precise understanding of the primordial BAO, and their direct connection to the sound horizon,%
\begin{marginnote}[]\entry{Sound horizon scale}{Distance the baryonic acoustic oscillation waves have traveled through the primordial plasma when baryons and photons stop interacting.}\end{marginnote}%
which we review in supplemental material \cref{ssec:primordial_oscillations}. 

The physical scale (length) of the sound horizon in the early Universe is a quantity that can be known or calibrated within the adopted model of the early Universe with high precision; it can thus be adopted as a {\it standard ruler}.%
\begin{marginnote}[] \entry{Standard ruler}{Object or feature of a known length. It can be used to measure distances when its angular size is observed.}\end{marginnote}%
The sound horizon leaves specific imprints in the observable Universe,  such that the angular sound horizon scale is then identified  and measured in observations and  in particular in the BAO signature on large-scale structure. Once calibrated, this early-Universe standard ruler can play a role similar to that of the local anchors and calibrate the \enquote{inverse distance ladder}\citep{Aubourgetal15,Cuesta:2014asa}.

\subsection{CMB anistropies}\label{ssec:CMBmain}

The CMB anisotropies (both temperature and polarization) have played a fundamental role in establishing precise constraints on the Hubble parameter and the establishment of the Hubble tension. This is presented in supplemental material \cref{ssec:CMB_SM}.

The final constraints on the cosmological parameters are usually derived using global parameter fits, whereby a cosmological model for a given set of parameters is used to obtain a prediction for the entire angular power spectrum which is compared to the observations. In this way, to each set of parameters a likelihood value can be assigned. Typically the Bayesian posteriors of a parameter inference run for a given model and a given set of parameter priors are reported. However, while this approach might initially seem to obscure the way in which $H_0$ can be extracted, analytical arguments can quickly demonstrate how the CMB can measure the \enquote{standard ruler} of the sound horizon, and anchor the Hubble parameter, which we present in Supplemental material \cref{ssec:CMB_physintyerp}. 
Even in $\Lambda$CDM, there are many degeneracies preventing directly inferring $h$ from the sound horizon angle, however  the CMB is subject to various additional effects that  do depend on the Hubble constant.
While  the specific physical signature for $h$ in the CMB is a bit tricky to pin down, and it is not yet decisively clear if there is one or multiple $h$ that can be measured in the angular spectrum, what is already clear is that the determination of the CMB has extremely small uncertainty, with
\begin{equation}
    [2.a] H_0 = 67.36 \pm 0.54 \,\hunit
\end{equation}
from \cite{PlanckParams2020} using {\color{red} the full dataset  comprising } Planck temperature, polarization, and lensing spectra analyzed in the $\Lambda$CDM model. The reliance on these multiple effects to break the underlying degeneracies causes a non-negligible dependence of the inferred Hubble parameter on the underlying cosmological model used in the analysis. For example, in a curved Universe, the Hubble parameter value inferred can be significantly lower since the curvature rescales the angular diameter distance of \cref{eq:angular_diameter}. The same is true for a Universe with non-trivial  dark energy dynamics or evolution.
In particular, 
CMB constraints on $H_0$ rely on assumptions about both the early-time physics as well as the evolution of the Universe from recombination to $z=0$. The assumption of a cosmological model (especially the $\Lambda$CDM model) very rigidly connects the quantities between these epochs.

The reliance on the multitude of additional effects and the resulting model dependence can be reduced if external calibrations of the relative expansion history are used, such as through BAO (see \cref{sec:bao} below) or using uncalibrated SNeIa, since then the measurement of the physical energy densities in the CMB more directly determines the Hubble constant. It is because of this that  late-time solutions rescaling  the angular diameter distance of \cref{eq:angular_diameter_delta} are generally much more restricted once these data are added (see also \cref{ssec:graveyard}).

\subsubsection{The many flavors of  $H_0$ measurements from CMB}
The number reported above is the \enquote{consensus} value: a variety of  variations (data choices, data cuts, analysis choices, etc.) have been presented which serve to asses the robustness of the consensus value. We discuss a representative sample.

The final angular power spectra derived by the \cite{PlanckParams2020} analysis show several peculiar features detailed in \cite{PlanckShifts}. In particular, there is a  shift between the parameters obtained from different sources (polarization or temperature) and between different multipole ranges ($\ell<800$ or $\ell>800$), though at this point the differences are small enough to still be explained by statistical effects.\footnote{There are also large-angle anomalies, already known since the mid-2000s. See  \cite{Copi:2010na} for a review.} The split for $\ell <800$ corresponds to the range where also the earlier WMAP experiment has measured the CMB anisotropies in \cite{WMAP9yr}, with excellent agreement between the two experiments.

\begin{table}[]
    \centering
    \begin{tabular}{| c l|c|}
    \hline
   Experiment & Dataset & $H_0$[km/s/Mpc] \\
        \hline
        \hline
        Planck & TT+TE+EE (+lensing) & $67.36 \pm 0.54$ \\
        Planck & TT+TE+EE & $67.27 \pm 0.60$ \\
        Planck & TT+TE+EE ($+\Omega_k$) & $54.4^{+3.3}_{-4.0}$\\
        Planck & TT+TE+EE ($+N_\mathrm{eff}$) & $66.4 \pm 1.4$\\
        Planck & TT+TE+EE ($+A_L$) & $68.28 \pm 0.72$\\
        Planck & TT+$\tau_\mathrm{reio}{}^{\rm a}$ ($+N_\mathrm{eff}$) & $66.5 \pm 2.3$\\
        Planck & TT+$\tau_\mathrm{reio}{}^{\rm a}$ & $66.88 \pm 0.92$  \\
        Planck & TT $(\ell <800)$+$\tau_\mathrm{reio}{}^{\rm a}$  & $70.0 \pm 1.9$ \\
        Planck & TT $(\ell >800)$+$\tau_\mathrm{reio}{}^{\rm a}$ & $64.2 \pm 1.3$ \\
        Planck & EE & $69.9 \pm 2.7$ \\
        \hline
        [2.$d$] WMAP & TT+TE+EE & $70.0 \pm 2.2$ \\
         ACT & TT+TE+EE & $67.9 \pm 1.5$ \\
         SPT & TT+TE+EE & $68.3 \pm 1.5$ \\
         \hline
         ACT+WMAP & TT+TE+EE & $67.6 \pm 1.1$ \\
         SPT+WMAP & TT+TE+EE & $68.2 \pm 1.1$ \\
         ACT + Planck & TT+TE+EE & $67.53 \pm 0.56$ \\
     SPT+Planck & TT+TE+EE & $67.24 \pm 0.54$ \\ 
     SPT+Planck+ACT & TT+TE+EE &    $67.49 \pm 0.53$\\
     \hline
    \end{tabular}
    \caption{
    Summary of constraints on $H_0$ from different CMB experiments. The experiments include Planck \citep{PlanckParams2020}, ACT \cite[Tab.~4]{ACTDR4}, SPT from \cite[Tab.~4]{SPT2022}, and WMAP from \cite[Tab.~3]{WMAP9yr}. TT refers to the CMB temperature anisotropies auto-correlation, EE  to the E-mode polarization auto-correlation, and TE  to the their cross-correlation. The brackets specify specialized analysis choices, such as a variation of the underlying model ($N_\mathrm{eff}$ is the effective number of neutrino species, $A_L$ a rescaling of the amplitude of the lensing power spectrum, and $\Omega_k$ the curvature parameter) or a restriction of the multipole range. When ACT and Planck are combined, their overlap in multipoles is removed.
    }
    \begin{tabnote}  
    ${}^{\rm a}$The optical depth (otherwise degenerate) in these cases is constrained by using the Planck polarization autocorrelation restricted to multipoles $\ell < 30$.
    \end{tabnote}
    \label{tab:cmb_h0}
\end{table}

\begin{figure}[h]
    \centering
    \includegraphics[width=\textwidth]{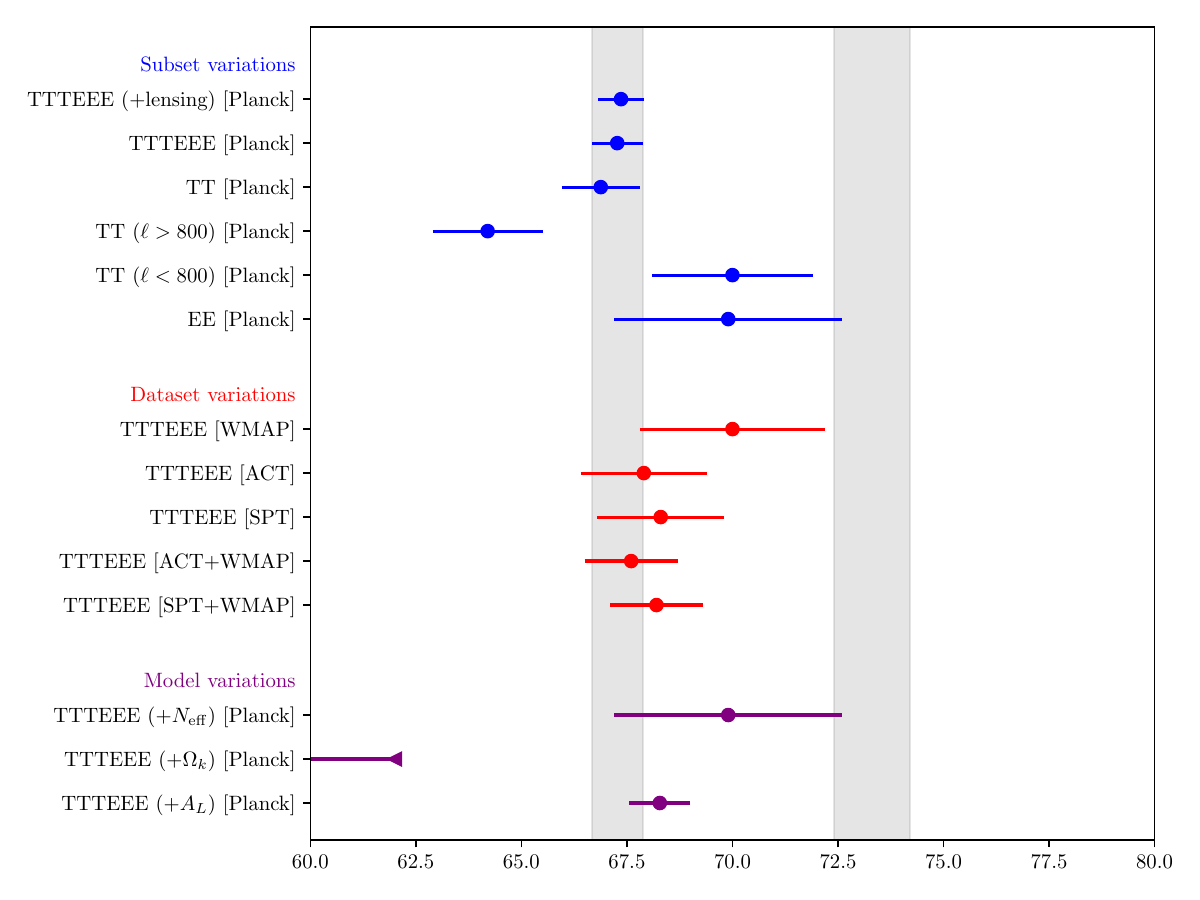}
    \caption{Measurements of $H_0$ from the CMB anisotropy power spectra summarized, see \cref{tab:cmb_h0} for further references.  Note that there is a scatter among the ``subset variations" which include very partial data, as discussed in \cite{PlanckShifts} the scatter is not inconsistent with  statistical noise. Since the mean value for the Planck TT+TE+EE ($+\Omega_k$) is too low, we only show the 95\% upper limit in this case. This case is known to present large parameter degeneracies and prior volume effects.}
    \label{fig:H0_whisker_cmb}
\end{figure}

Other ground-based experiments independently confirm the low Hubble constant. From the south pole telescope (SPT, \cite{SPT2022}) a value of $H_0 = 68.3 \pm 1.5\,\hunit$ [2.$b$] can be found, and from the Atacama cosmology telescope (ACT, \cite{ACTDR4}) a value of $H_0 = 67.9 \pm 1.5\,\hunit$ [2.$c$] can be found. Combinations with the large scale CMB anisotropy observations from the WMAP experiment decrease the uncertainties to $1.1\,\hunit$ for both experiments, while not shifting the mean value.

\Cref{tab:cmb_h0} represents a summary of these measurements, which is also graphically summarized in \cref{fig:H0_whisker_cmb}. 
The different experiments agree remarkably well, which is expected to some extent as they observe the same sky. However, it should be stressed that their instrumental systematics are fully independent, potential analysis systematics are also largely independent, and there is a decent check on possible systematics arising from foreground residuals because the frequency bands and the foregrounds treatments are not identical (although there is some degree of overlap in the foreground modeling). It is interesting to note that polarization-only determinations agree well with the temperature-based ones.

The  trend for higher central $H_0$ values when only multipoles $\ell < 800$ for the temperature are included and lower central values for $\ell>800$ (both with larger uncertainties) is clearly visible.

Indeed, with the $\ell<800$ results from temperature only ($H_0 = 70.0\pm 1.9\,\hunit$), the Hubble tension with respect to \cref{eq:H0cepheid_boosted} would only be at the level of $1.6\sigma$ (statistically consistent). However, the inclusion of  $\ell < 800$  temperature-polarization cross correlation (no lensing) yields $H_0 = 67.32 \pm 0.95\,\hunit$ ($4.6\sigma$).
 
Studies indicate that the $\ell < 800$ trend is consistent with being a mere statistical fluctuation, possibly related to the large-scale \enquote{CMB anomalies} (see \cite{PlanckShifts}).\footnote{Another such anomaly arises when allowing for an overall rescaling of the lensing power spectrum in the theoretical calculation through the parameter $A_L$\,. While the $\Lambda$CDM expectation is $A_L=1$, the analysis of \cite{PlanckParams2020} finds higher values of $A_L>1$. This is taken by some as another hint at a possible inconsistency of the model between describing the multipole ranges of $\ell < 800$ and $\ell>800$ (where lensing is more relevant), though the un-physicality of the parameter, the magnitude of the deviation ($\sim 2.8\sigma$), and the lack of such trends in other datasets (Planck lensing, SPT, and ACT all do not favor $A_L \neq 1$) could hint at a merely statistical effect (a \enquote{look-elsewhere} effect).}

To summarize, CMB observations  within the standard $\Lambda$CDM model,  all consistently define a low \enquote{camp} for $H_0$ of around $68\hunit$ (with a $\sim 1$\% error). What all these determinations have in common is the adopted cosmological model, and we show those popular simple modifications with the largest impact on the inference of $H_0$ from the CMB.

\subsection{Large-scale structure clustering}\label{sec:bao}

The reach of the CMB in terms of redshift can be extended by combining it  with  lower-redshift  probes that are calibrated using direct information from the CMB anisotropies. One such example is the calibration of the sound horizon of the BAO visible both in the CMB and the late clustering of large scale structure.

The BAO signal in galaxy clustering surveys can be seen either as an oscillatory signal in the power spectrum (see supplemental material \cref{ssec:primordial_oscillations}) or -- after a Fourier transform -- as a peak in the real-space correlation function: galaxies are more likely to be found a sound horizon apart than at other distances. For this reason, it is common to use  \enquote{feature}, \enquote{peak}, or \enquote{oscillations} interchangeably when referring to the BAO measurements.

There are two philosophies to  extract information about the cosmological parameters from galaxy redshift surveys. One approach selects a cosmological model with a certain number of adjustable parameters and then performs parameter inference by comparing (via a likelihood function) the model's theoretical prediction for the galaxy clustering observable (or selected statistics) with the measurements from the data
(we call this \enquote{full-modeling}). Of course, the full procedure of parameter inference  needs to be repeated when changing the  model. This approach extracts the largest amount of information from the observations (i.e. the smallest statistical errors around the model's parameters) but is more easily biased by un-modeled systematics. Instead, the second approach attempts to robustly compress the information contained in the observables into only a few physically interpretable summary statistics. 

The primary focus of this review  is on the compressed approach, since it is  easier to interpret physically, and has been shown to be model-agnostic (the compressed variables do not depend on the  cosmological model choice) but we will also report the results from the full-modeling approach. In particular, we  focus on the BAO scale as  observed in the clustering of galaxies, and how it can be used as an alternative probe of the Hubble constant once the  corresponding standard ruler is calibrated. The redshift space distortion signal at present does not contribute significantly to $H_0$ and is discussed in supplemental material \cref{ssec:RSD}. The same considerations apply, with only small changes,  to the clustering of other tracers (quasars and  Lyman-$\alpha$, see below).

\subsubsection{A short history of BAO measurements} 

The BAO feature in the matter power spectrum (as traced by the galaxies) is a few \%-level modulation  at large scales, hence the galaxy three-dimensional  distribution needs to be mapped over large  areas of the sky and large volumes to  gain enough statistical power to identify the BAO feature.
The BAO feature was first detected on galaxy clustering data in Sloan Digital Sky Survey (SDSS) luminous red galaxies~\citep{Eisenstein:2005su} and in 2-Degree Field Galaxy Redshift Survey (\mbox{2dFGRS},~\cite{cole05}), and first employed as a robust cosmological probe a few years later in~\cite{percivaletal:2007}. The sensitivity in detecting the BAO peak in a galaxy sample is proportional to the number of modes sampled (i.e., the size of the observed volume  but also  the galaxy density of the observed sample). Better BAO measurements at increasing redshift yield  improved constraints on the  expansion rate $E(z)$ but require ever larger survey volumes being sampled by a high density of objects with precise redshift determinations. This has been a major driver of observational efforts over the last decade.
 Spectroscopic galaxy samples have been designed with this specific purpose: the SDSS Main Galaxy Sample (MGS,~\cite{Ross2015_MGS}), WiggleZ~\citep{WiggleZ}, SDSSIII-Baryon Oscillation Spectroscopic Survey (BOSS,~\cite{alam_clustering_2017}), and SDSSIV-extended BOSS (eBOSS,~\cite{Alam2021}) surveys have exploited the BAO feature in the galaxy and quasar clustering data, reaching in some cases precision of around $1\%$ in the determination of the peak position. 

The so-called {\it reconstruction} technique has greatly helped to reach that notable milestone. This technique  enhances the significance of the BAO feature by removing the effect of  the peculiar velocity bulk flows that tend to smear out the BAO feature~\citep{padmanabhan_reconstructing_2009}. This is achieved by displacing the observed galaxies' positions backwards in time using the Zel’dovich approximation; the displacement is  calculated from the observed local density field. 
In other words, while the power spectrum does not capture phase information (after all in a Gaussian random field phases are random, and linear gravitational evolution does not couple modes and therefore does not change the phases) mildly non-linear evolution does couple phases: this phase coupling  broadens the BAO peak. This gravity-induced  phase information is however  present in the overdensity map, and this is what is extracted and employed in reconstruction to \enquote{unblur} the BAO peak. This step relies on Newtonian dynamics as the scales involved are not large enough to warrant worrying about relativistic corrections.

In addition to the galaxy and quasar clustering BAO, the detection of the BAO peak using the Lyman-$\alpha$ forest technique has allowed sampling the expansion history back to epochs where no galaxies or quasars can be efficiently identified~\citep{dumasdesBorboux2020}. By observing the absorption lines from background quasars the Lyman-$\alpha$ technique is able to detect density fluctuations of neutral hydrogen intergalactic clouds, which also have the BAO feature imprinted. All these efforts have culminated with a sampling $E(z)$ along 10 billion years of cosmic evolution, from low redshifts of $z\sim0.1$ using bright nearby galaxies up to high redshifts of about $z \sim 3.5$ using Lyman-$\alpha$ forest lines. This effort has yet not reached its cosmic variance limit, which is now one of the drivers of  ongoing surveys such as the Dark Energy Spectroscopic Instrument (DESI,
~\cite{aghamousa_desi_2016,DESI_DR02}) or Euclid ~\citep{laureijs_euclid_2011}.

\subsubsection{Uncalibrated BAO}

The observed BAO peak position in the distribution of galaxies or gas in the late universe (or any time after the drag epoch) is determined by two effects: 1) The actual comoving size of the sound horizon at drag epoch $r_d$\,, giving the initial position of the peak (though it is slightly modified by non-linear evolution) and 2) the mapping from comoving space to observed redshifts and angles, which is intimately tied to the expansion history of the universe at the redshift of observation as well as the integrated expansion history until today.
A model-agnostic analysis of the BAO feature in the three-dimensional maps of galaxies allows us to determine two BAO features: one angular feature across the line of sight, and one radial feature along the line of sight. Although the BAO clustering is isotropic in comoving space, the different conversions to observations with radial redshifts and perpendicular angular sizes motivate to measure two different quantities related to $H_0$
\begin{align}
    &\delta z =\frac{r_H}{r_d} & \,\,\,\,\,{\rm line\,\,of\,\, sight\,,\,\, with\,\,\,\,\,}& r_H(z)= \frac{1}{H_0}\frac{c}{E(z)}\label{eq:hubble_diameter}\\
     &\delta \vartheta= \frac{r_d}{r_A} &\,\,\,\,\, {\rm angular\,,\,\, with\,\,\,\,\,}&r_A(z) = \frac{1}{H_0}(1+z) \int_0^z\frac{c \, \mathrm{d}z'}{E(z')}\label{eq:angular_diameter_delta}
\end{align}
Note that only the ratios of length scales can be measured as there is no immediate information about absolute distances. Indeed, both ratios have a perfect degeneracy between $r_d$ and $H_0$ if no additional information is supplied. However, the uncalibrated expansion history $E(z)$ can still be measured from the BAO. In particular, the ratio $\delta z/\delta \vartheta \propto r_H/r_A$ is a direct tracer of $E(z)$ and doesn't need to be calibrated. It is directly related to the Alcock-Paczy\'nski effect, since in the correct cosmology the transversal and radial measurements of the BAO sound horizon should agree (see \cite{1979Natur.281..358A}). Similarly, the BAO measured from different redshifts can be combined to probe $E(z)$ at different redshifts.
The observed absolute values of the quantities $\delta z$ and $\delta \theta$ instead constrain the combination~$hr_d$\,.
In summary, these various \enquote{uncalibrated} BAO measurements can be combined to measure $E(z)$ or equivalently $\Omega_m$  and  $h r_d$\,, but the Hubble constant itself remains elusive.

\subsubsection{Calibrating the BAO}
The uncalibrated BAO does not give a measurement of the Hubble constant $H_0$\,. 
In order to calibrate (or anchor) the BAO and to deliver a measurement of $H_0$ or $H(z)$ we need to set the absolute scale of $r_d$.

The length of the sound horizon as a standard ruler can be obtained in more than one way.
It can be extracted from the CMB either using the full posterior of $r_\mathrm{d}$ (which is not independent of the $H_0$ posterior of the CMB, see \cref{ssec:bao_cmb}) or using only summary information (see \cref{ssec:bao_cmb_priors}). It can also be  calibrated using other measurements of physical energy densities such as BBN (see \cref{ssec:bao_bbn}), or even different features in the galaxy power spectrum (see \cref{ssec:shapefit}).

The BAO comoving scale in a $\Lambda$CDM model depends on the physical densities as (see \cite{2hpaper})
\begin{equation}
\label{eq:rw_lcdm} r_{\rm d}=\int^{z_d}_{\infty} \frac{c_s(z)}{H(z)}dz \simeq \mid_{\Lambda {\rm CDM}} \frac{147.05}{\rm Mpc} \left(\frac{\Omega_m h^2}{0.1432}\right)^{-0.23}\left(\frac{N_{\rm eff}}{3.04}\right)^{-0.1} \left(\frac{\Omega_{\rm b} h^2}{0.02236}\right)^{-0.13}~,
\end{equation}
where $N_{\rm eff}$ denotes the number of effective neutrino species. 
It should therefore be clear that the length of the standard ruler as observed in the late time Universe, is calibrated on early-time physics. There is broad consensus in recognizing that this physics constitutes the most robust part of the (standard) cosmological model.

\subsubsection{BAO combined with full CMB anistropies}\label{ssec:bao_cmb}

A direct combination of the full anisotropy constraints from Planck (including CMB lensing) with the BAO from SDSS BOSS gives $H_0 = 67.66 \pm 0.42\,\hunit$ \citep{PlanckParams2020};  the more recent results from BOSS+eBOSS with Planck (without lensing) are quite similar at $H_0=67.61 \pm 0.44\hunit$ \citep{Alam2021}. As mentioned in \cref{ssec:CMBmain}, the primary advantage of combining the CMB with the BAO in this way is reducing the residual parameter degeneracies  in the CMB-determination of $H_0$\,.

\subsubsection{BAO combined with CMB priors}\label{ssec:bao_cmb_priors}

To offer insight into what part of the CMB spectrum (and thus what aspects of early-time physics) or what possible systematics (if any) might be driving the signal, it is useful to look at how the $H_0$ constrain changes if only summary information is retained.
This summary information can be  built so that it can be made  more robust to systematic issues.
The  most common choice is the $r_d$ information, as it is insensitive to late-time physics assumptions  (e.g., \cite{Verde17}).

As an example, the eBOSS+BOSS BAO analysis of \cite{Alam2021} yields a high precision measurement of $h r_d = 100.4 \pm 1.3 \,\inversempc$, which together with a prior of $r_d =  147.09 \pm 0.26$ from \cite{PlanckParams2020} yields $H_0 = 68.28\pm 0.89\,\hunit$.

The addition of SNeIa to the CMB+BAO  combination  very tightly constrains the expansion rate $E(z)$, which converts the BAO into an almost direct measurement of $h r_d$ at very high precision, even for a very flexible parameterization of the expansion history. For example, \cite{Camarena2020} use $r_d = 147.09\pm 0.26$Mpc and find $H_0 = 69.71 \pm 1.28\,\hunit$ in a model-independent analysis using Pantheon supernovae and BOSS BAO (only information from $\delta \vartheta$ at $z=0.38,z=0.61$), while \cite{Macaulay2019} use $r_d = 147.05 \pm 0.30$ Mpc and find $H_0 = 67.8\pm 1.3\,\hunit$ using BOSS BAO and various alternative SNeIa samples.

\subsubsection{Bypassing CMB data altogether: BAO combined with BBN}\label{ssec:bao_bbn}
The  sound horizon standard ruler can be calibrated in a way that is completely independent of  CMB observations. This is done through the so-called Big Bang Nucleosynthesis (BBN) constraint on $\Omega_b h^2$ (and $N_{\rm eff}$). 
We have established that the unnormalized BAO  constrain $\Omega_m$ and $r_dh$ which, from \cref{eq:rw_lcdm},  scales like $r_dh\propto \Omega_m^{-0.23}h^{0.52}N_{\rm eff}^{-0.1}(\Omega_b h^2)^{-0.13}$.

In fact, the primordial abundances of light elements are tightly connected to the baryon-to-photon ratio (effectively determining $\Omega_b h^2$ for a fixed\footnote{The COBE/FIRAS measurement of the CMB temperature has such small error bars associated that the CMB temperature can be considered fixed for this application.} CMB temperature) as well as the early time expansion rate (determined through any additional relativistic dark species at BBN for a fixed CMB temperature). Both can be directly constrained through the  late-time measurement of the deuterium and Helium abundances obtained from the analysis of absorption features in spectra of quasars and emissions of HII regions of metal-poor galaxies. Measurements of $\Omega_b h^2$ from BBN allow the BAO to self-calibrate (using the derived $\Omega_m$) and thus to deliver the measurement of the only remaining parameter: $h$.

It is interesting to note that this approach is only sensitive to the evolution of background quantities, while the CMB-based calibration of $r_d$ is based on a combination of both  the  background and  the perturbation  evolution.

We reiterate that  this calibration still relies on {\it i}) early-Universe information through $\Omega_b h^2$ from BBN, and {\it ii}) the assumption of a model for  early-time physics (usually the $\Lambda$CDM model is adopted, see \cref{eq:rw_lcdm})  in order to derive the scale of $r_d$\,. Only when both are adopted the determination of $h$ from BAO becomes possible. As such, the BAO+BBN and CMB anisotropy probes are sensitive to very different systematics in terms of instrumentation, observations, measurement methodology, etc., but they share assumptions about the early universe model for the baryon-photon plasma. A recent analysis of the BAO+BBN probe from \cite{Schoneberg2022} finds $H_0=67.6 \pm 1.0\,\hunit$ [2.$e$].

\subsubsection{Bypassing CMB data altogether: BAO combined with ShapeFit}
\label{ssec:shapefit}
It is well known that the clustering of galaxies carries more information than just the baryonic acoustic oscillations imprinted in its power spectrum. In particular, the power spectrum of the galaxy clustering is subject to a turn-over at matter-radiation equality, baryonic suppression carries information from BAO amplitude, phase, and damping, and is subject to non-linear structure formation which couples different Fourier modes and enhances small scales power. In addition, redshift space distortions induced by peculiar velocities introduce clustering anisotropies along the line of sight which carry information about the growth of cosmic structures. Of all these effects, it has been recently shown that the large-scale shape of the matter power spectrum
can be used as an effective probe of the physical matter density $\Omega_m h^2$  when interpreted within the $\Lambda$CDM model (or a well-defined set of assumptions which are also in common with the $\Lambda$CDM ones, see \cite{2hpaper} for a discussion), once the physical baryon density $\Omega_b h^2$ is known \citep{ShapeFit}. With the matter density parameter $\Omega_m$ constrained by the expansion history as measured from the uncalibrated BAO (or even from uncalibrated cosmological high redshift SNeIa) a novel determination of $h$ can be made by combining these constraints.

The \enquote{ShapeFit} approach of \cite{ShapeFit}  
involves measuring the power spectrum slope at a given wavenumber (the \enquote{Shape}), which depends on the details of perturbations growth around matter radiation equality, the baryonic suppression (i.e., the shape of the knee of the matter transfer function), and the primordial tilt of the power spectrum $n_s$\,.  Thus $h$ can be determined  given  priors on $\Omega_b h^2$ and $n_s$.

Using the BOSS+eBOSS luminous red galaxies, quasars and Lyman-$\alpha$ samples covering the $0.2<z<3.5$ range  $H_0 = 68.16 \pm 0.67\,\hunit$  is obtained for a fixed value of $n_s=0.97$ and a simplified BBN prior of $\Omega_b h^2 = 0.02235 \pm  0.00037$ \citep{Brieden:2022lsd}, or
$H_0 = 68.3 \pm 0.7\,\hunit$
when using an $n_s = 0.9637 \pm 0.0044$ prior and $\Omega_b h^2$ consistently derived from BBN measurements using updated theoretical predictions \citep{Schoneberg2022}. Even for a relatively broad Gaussian prior of $\sigma_{n_s}=0.04$  on $n_s$ the derived constraints remain unchanged \citep{2hpaper}.

\subsubsection{Bypassing CMB data altogether:  full-modeling of the power spectrum}
\label{sec:fullmodeling}
In the {\it full modeling} approach, direct fits to the power spectrum  constrain the free parameters of the  adopted model, including $H_0$\,. Usually, these fits need to be supplemented by a BBN prior for the baryon abundance $\Omega_b h^2$. Although these are not pure BAO+BBN fits, they employ the BAO information, but also other features, such as the shape of the matter power spectrum (used by the ShapeFit approach), and  the phase and the damping of the BAO features (not used by the ShapeFit approach).
Interestingly full modeling  analyses do not use a $n_s$ prior (since that is measured self-consistently from  the power spectrum).

It has been shown that both full modeling and the ShapeFit approach deliver very similar constraints, in particular for the determination of $H_0$\,, at least for the usual $\Lambda$CDM model extensions \citep{Mausetalinprep}. Unlike the ShapeFit approach, the full modeling approach needs to assume a model {\it before} the fitting procedure. Instead, ShapeFit assumes the model {\it after} the fitting procedure. In the last reported studies of full modeling analyses, \cite{Zhang2022} find $H_0=68.19\pm0.99\,\hunit$ when fitting to the correlation function of the BOSS luminous red galaxy sample along with the eBOSS quasar sample, and also using the BAO information from the Lyman-$\alpha$ forest as well as a BBN-motivated prior. Similarly, \cite{philcoxetal22} find $H_0=69.3^{+1.1}_{-1.3} \,\hunit$ when fitting the BOSS luminous red galaxy sample.\footnote{Note that both \cite{Zhang2022} and \cite{philcoxetal22} include also uncalibrated BAO information extracted from the post-reconstructed catalogues in addition to the direct fits of the pre-reconstructed catalogues.} See \cite{Simon2023, BrinchHolm23} for a detailed comparison of both approaches.

\subsubsection{Bypassing CMB altogether: Sound-horizon independent information}\label{ssec:equality_scale}
Early-time physics provides another anchor, namely the matter-radiation equality scale, which can be used in several ways to extract a CMB-independent $H_0$ constraint. While the relevant physical processes setting the sound horizon scale happen at $z\sim 1100$, the matter-radiation equality occurs at $z\sim 3300$. There are at present three different approaches to calibrate the inverse distance ladder on this anchor. 
\paragraph{ShapeFit based} In \cite{2hpaper} an uncalibrated BAO measurement (constraining mostly $\Omega_m$) is combined with the information from the Shape (indirectly measuring $\Omega_m h^2$), leading to a constraint of $H_0 = 70.2^{+1.9}_{-2.1}\,\hunit$, even without the sound-horizon calibration of the BAO. This signals that the assumed modeling of the sound horizon is unlikely to be at fault alone for the discrepancy between the different values of the Hubble constant from \cref{sec:distance,sec:invdist}.

\paragraph{Marginalizing over the sound horizon}
In the {\it full modeling}  approach, instead of using compressed variables and uncalibrated BAO, it is possible to marginalize over the sound horizon information explicitly (in order to keep only the non-$r_s$ information). In such  analyses (while including a BBN-motivated prior on $\Omega_b h^2$), \cite{Farren2022} find $H_0 = 69.5^{+3.0}_{-3.5}\,\hunit$ from older BOSS power spectrum data, and \cite{Philcox2022} find $H_0 = 67.1^{+2.5}_{-2.9}\,\hunit$ from BOSS combined with CMB lensing and Pantheon+ SNeIa. 

\paragraph{BAO combined with $P(k)$ turnover}
In a first indicative study \cite{Bahr-Kalus2023} use the actual position of the turnover in the power spectrum measured by eBOSS quasar sample as a standard ruler. The turnover scale marks matter-radiation equality and is therefore set by $\Omega_m h^2$ (with some weak dependence on $\Omega_b h^2$ as the baryon-suppression begins on the same scales). Using either SNeIa or uncalibrated BAO to determine $\Omega_m$, the Hubble constant can be extracted, and they find $H_0 =63.3^{+8.2}_{-6.9}\,\hunit$. This measurement is completely independent of any sound horizon information and likely also very independent of baryonic physics in the early universe. While the uncertainty is currently too high to be competitive with other methods, this technique will become increasingly relevant  with future and ongoing large scale structure surveys.

\subsection{Putting it all together: some thoughts on systematics and risk assessment.}\label{ssec:cmb_systematics}
It is clear that the  $H_0$ determination of the inverse distance ladder (either extrapolating a one-rung ladder using only CMB data within a model, or in a two-rung ladder calibrating the sound horizon and then using BAOs) relies on two components: the anchor and the expansion history. Data put constraints on both, but there is a residual model-dependence. It is therefore natural to ask:  where could systematics (or unknown unknowns) hide?
It is clear that experimental/instrumental systematics  large enough to shift the $H_0$ inferred within $\Lambda$CDM  cannot hide in CMB data: the $H_0$ determination does not change significantly when  using different CMB experiments (as long as the damping tail is included) and when excluding CMB data altogether. However, the BAO sound horizon anchor and the equality scale anchor both depend on the same assumed early-time physics as the CMB. 
\subsubsection{Implications for early-time physics}
A shift in $H_0$ can be obtained if the dependence of 
 $r_d$ on the $\Lambda$CDM parameters  were to be  different than expected.
This requires introducing some new physics, which -- as \cite{Knox_Millea20} highlight -- need to be dynamically  relevant in the e-fold\begin{marginnote}[]
\entry{e-fold}{Timespan required for the expansion of distances within the universe by a factor of $e \approx 2.7183$.}%
\end{marginnote}%
of expansion before recombination  and disappear quickly at recombination, in order not to spoil the  excellent agreement of theory and data. It turns out that this is very hard to do in practice, with physically motivated or even toy models, as most modifications break havoc, typically on the CMB damping tail. Of course, this does not prove that such models do not exist, but it makes model-building very constrained. See the related discussion in \cite{Kamion_Riess22} and \cref{ssec:graveyard}.
However, changes in pre-recombination physics that shift $r_s$ by $\sim 7\%$ need not introduce a tension between the two early Universe anchors of $r_d$ and the equality scale.

\subsubsection{Implications for the shape of the expansion history}\label{sec:expansionhistory}
One of the reasons behind the success of the $\Lambda$CDM model is its power of predictability: predictability means that the model's predictions (and thus the model itself)  can be falsified by  observations. When instead new observations confirm the predictions this is, justifiably so, taken as a conformation of the model.

\begin{figure}[h]
\begin{minipage}[c]{0.5\linewidth}
\centering
\includegraphics[width=1.9in, trim = 6cm 0cm 9.8cm 0cm, clip]{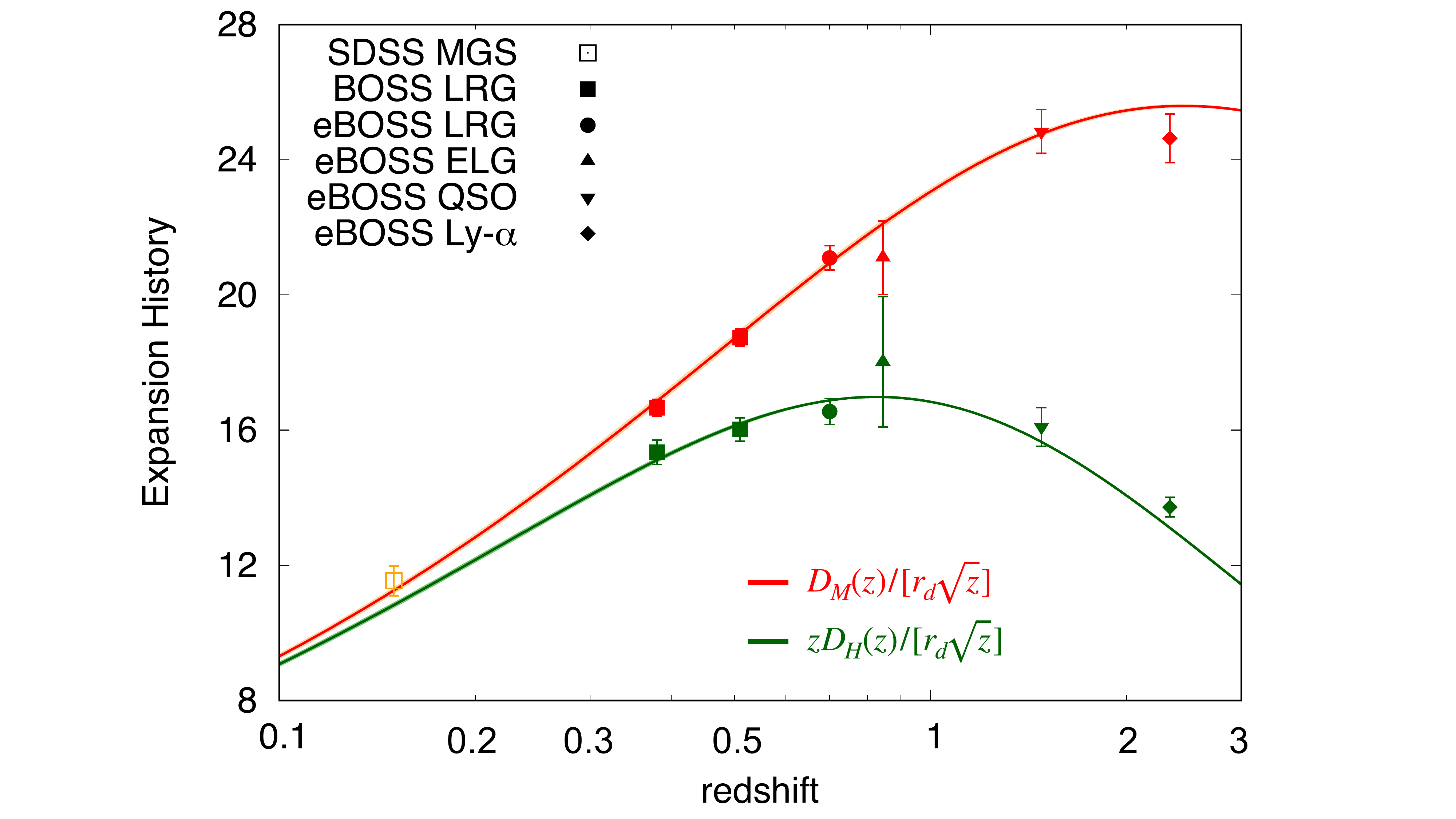}
\end{minipage}
\begin{minipage}[c]{0.5\linewidth}
\centering
\includegraphics[width=2.5in, trim = 0.3cm 0cm 0cm 0cm, clip]{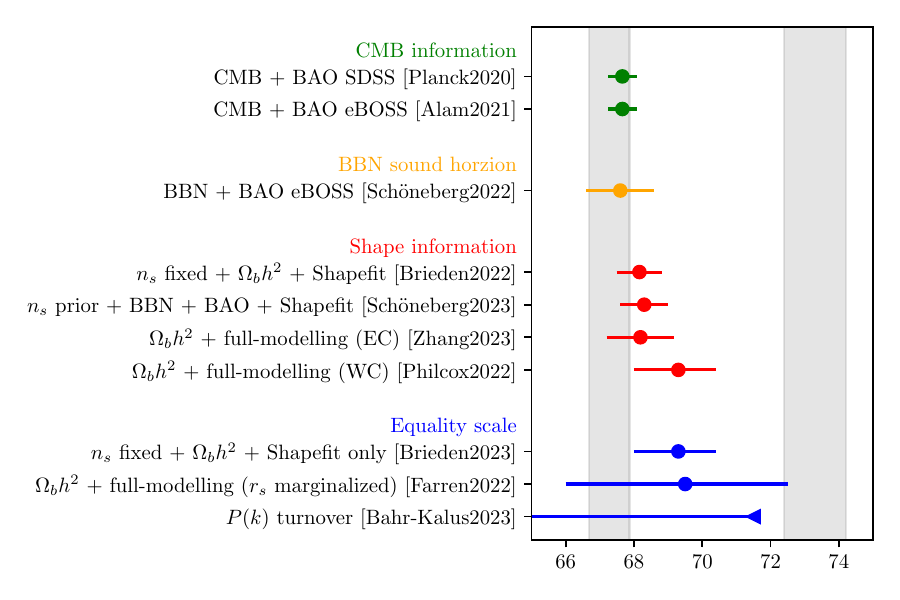}
\end{minipage}
\caption{Left: BAO measurements (tracing the expansion history) as a function of redshift. Red points correspond to transverse BAO, while green points to radial BAO.  The points do not depend on the assumed value of $H_0$. The red and the green curves are not fit to the data: they are the Planck {\bf bestfit predictions} for the standard cosmological model \citep{Alam2021};  as such in a standard $\Lambda$CDM model, it implies  an $H_0$ value fully representative of the 68 camp. Ignoring CMB observations, as long as $r_d \, h\simeq 100$ Mpc and $\Omega_m\simeq 0.3$, a value of $H_0=73$ \hunit{}\ (or any other values) would still yield the predictions indicated by the solid lines (but not providing a good fit to the CMB and BBN data). Right: summary of $H_0$ determinations of Sec.~\ref{sec:bao}.}
\label{fig:expansionhistory}
\end{figure}

\Cref{fig:expansionhistory} (see also Fig 1 of  \cite{Alam2021}) illustrates the predictive power of the standard cosmological model and the  confirmation of its predictions by independent observations. Here a $\Lambda$CDM model calibrated on CMB (Planck) data and early-time physical processes predicts key quantities observed in the later Universe. 
It is also remarkable to note how well the measured  $hr_d$ from eBOSS ($h r_d = (100.4 \pm 1.3 ) \, \inversempc$, see \cite{Alam2021}) agrees with the Planck predicted value in $\Lambda$CDM of $h r_d = (99.1\pm 0.9) \, \inversempc$  \citep{PlanckParams2020}.

It is reasonable to ask  whether the expansion history of the Universe between recombination at $z\sim 1100$ and now $z\sim 0$ could hide a step or a feature which could reconcile the tension.  
Model-independent (or rather model agnostic)  reconstructions of the expansion history  (e.g., \cite{BernalTriangles, Camarena2020, Macaulay2019, Favale23} and references therein) show that despite the increased freedom of such parameterizations of $E(z)$, a feature that causes such a 7\% mismatch in $H_0$ is heavily disfavoured by the data.
In summary, $E(z)$ is measured empirically to be that of a standard (CMB-calibrated) $\Lambda$CDM model with only very little remaining freedom (which is reduced even further by the probes discussed in \cref{ssec:ages}).

\section{ABANDONING THE DISTANCE LADDER}\label{sec:noladder}
The approaches outlined here provide a one-step measurement of the Hubble constant, completely independent of the local distance ladder or the inverse distance ladder. The dependence on other cosmological parameters is also weak for these approaches.  The most interesting constraints arise from strong lensing time delays, standard siren measurements, and through the use of cosmic ages. The use of megamasers could also be argued to be a one-step measurement. It has been discussed in \cref{sssec:single_rung} as the best-measured maser distance is used as an anchor for the  distance ladder. 

\subsection{Lensing time delays} \label{ssec:timedelays}
Measurements of relative arrival times of multiply gravitationally lensed sources can provide an absolute distance scale, and this is what time-delay cosmography is about. 
Originally proposed by \cite{Refsdal1964} for multiply lensed transient phenomena (say a supernovae), the best cosmological constraints to date come from quasars (which show intrinsic variability, and remain bright for many years of observation, unlike supernovae) which are strongly lensed by galaxies into multiple images.
When the intensity of a strongly lensed background source varies over time, such as an active galactic nucleus (AGN), the variability pattern  in each of the multiple images is delayed in time due to the different light paths of the different images. The relative arrival time  between two images (say A and B) originated from the same source (O) is proportional to  the time delay distance $D_{\Delta t}$ and the difference of the Fermat potential between the two images $\Delta \tau$. 
\begin{equation}
{\Delta t}= D_{\Delta t}/c  \Delta \tau~.
\label{eq:timedelaydistance}
\end{equation}
The Fermat potential is in turn  a combination of  the geometric path difference and the local spacetime dilation (Shapiro delay). The time delay distance is  an absolute physical distance  and is given by 
$D_{\Delta t}= (1+z_d) D_dD_s/D_{ds}$, where the subscript  $d$ stands for the lens (deflector), $s$ for the source, $D_{s,d}$  denotes the angular diameter distance, and $D_{ds}$ is the angular diameter distance between the source and the lens. Constraints on the Fermat potential difference $\Delta \tau$ (coming from a given model of the system) and a measured relative time delay $\Delta t$ between two images of the same source can be turned into constraints of the time-delay distance $D_{\Delta t}$\,. This is what  the time-delay methodology yields.

As a physical distance, it is clear that $D_{\Delta t}$ is inversely proportional to $H_0$. While there is a mild dependence on the relative expansion history between $z=0$ and the redshift of the lens and the source, the conversion of time-delay  measurements to a constraint on the Hubble constant  is typically relatively stable with cosmology \citep{Taubenberger19,Wongetal20}.
While the time delay between images is a direct observable, the relative Fermat potential is not, and must be inferred, modeled, or constrained from other observations. 
Multiple imaged sources and their extended distortions in the lensed images  and/or the actually pixelated lensed images of extended sources (QSO or SNe host galaxies) can be used to constrain the relative Fermat potential, but there are degeneracies that imaging data cannot break. The most important is the mass-sheet degeneracy (MSD):\begin{marginnote}[] \entry{MSD}{Mass sheet degeneracy. Internal degeneracy involving mass distribution and geometry of the time delay lensing system.}
\end{marginnote}%
 changes to the mass distribution, which involve adding a uniform mass sheet to the lens plane while at the same time moving the source position (and shape) and therefore the distance, can change $D_{\Delta t}$ and $\Delta \tau$ coherently as to  leave the time delays $\Delta t$ unchanged. The MSD is a multiplicative transform of the lens equation which preserves image positions and relative lensing observables. 

It turns out that \cref{eq:timedelaydistance} including the MSD can be written as 
\begin{equation}
{\Delta t}=\lambda D_{\Delta t}/c  \Delta \tau
\label{eq:MSD}
\end{equation}
where $\lambda$ is the MSD parameter. It can be further broken into $\lambda=(1-\kappa_\mathrm{ext}) \lambda_\mathrm{int}$\,, where $\kappa_\mathrm{ext}$ is interpreted as the effective external convergence made up of all line-of-sight lensing contributions  focusing or de-focusing of the light rays, and $\lambda_\mathrm{int}$ is instead related to the radial density profile of the main lens.  
\cref{eq:MSD} is a purely mathematical degeneracy, which is why  physical priors on $\Delta \tau$ are needed to break such degeneracy. These may come either from other galaxies that are assumed to be physically similar to  lenses  or from the dynamics of the lens itself. 

Thus besides measuring time delays between at least one multiple-image pair, and spectroscopic redshifts of the source and the lens, a lens mass model and constraints on external lensing convergence are needed. 
This implies that one has to rely on non-lensing information that specifies and constrains the radial mass density profile  and/or  priors about the functional form of the radial mass density  profile.
The line-of-sight lensing contribution, $\kappa_\mathrm{ext}\,$, is then constrained by observations of tracers of large-scale structure or weak lensing measurements, which today achieve a precision of a few percent per sight line.   

To date, modeling of the deflector mass density profile represents the biggest contribution to the systematic uncertainty on the distance and therefore $H_0$\,. One possibility to constrain it arises from spatially resolved measurements of the  velocity dispersion of the lensing galaxy, which would be non-trivially changed by the MSD. By comparing the observed 2D velocity dispersion with the predictions based on the mass sheet degeneracy, it's possible to break the MSD:  observations of  stellar kinematics in the lens galaxy can be used very efficiently to constrain the mass distribution of the lens in a way that is highly complementary to lensing constraints alone.

In terms of results,  following the original idea of  Refsdal, an analysis of supernova Refsdal lensed by a cluster of galaxies, which first appeared is 2014 and reappeared in 2015 (as predicted by some of  the lens models), has recently yielded 
$H_0 = 64.8\pm 4.3\,\hunit$, for which the posterior probabilities of various mass models have been added in a weighted way, rather than being treated as systematic uncertainty. It should be noted that the different adopted models predict mutually incompatible time delay differences.

While there are several lensed supernovae discovered in the past few years,  most of them do not have sufficient time-delay measurements for cosmology.  One system though, SN H0pe, that is  lensed by a galaxy cluster and recently discovered in JWST imaging, is promising to yield a competitive distance measurement (see \cite{Fryeetal23}).

To date, most time-delay measurements have been made with quasars. The COSMOGRAIL, H0LiCOW, SHARP, and STRIDES collaborations (together TDCOSMO),\footnote{See \url{http://cosmograil.org}, \url{www.h0licow.org/}, \url{https://sites.google.com/view/sharpgravlens/}, \url{https://strides.astro.ucla.edu}, and \url{https://obswww.unige.ch/~lemon/tdcosmo-master/index.html}.} building on more than a decade of painstaking work (see for example \cite{Suyu2017,Wongetal20,Millon2020} and references therein), collected observations of 7 lensed quasar systems with multiple images, for which variability monitoring goes back up to two decades in some cases (see COSMOGRAIL).

The approach followed initially was to break the mass-sheet degeneracy by making standard assumptions about the mass density profile of the lens galaxy. In fact,  allowing a full free-form $\lambda$ assumes that we know nothing of galaxies.
Two descriptions for the deflector mass density profile were considered: either a power law or the superposition of the observed stellar component (assuming a constant mass-to-light ratio) and a standard Navarro-Frank-White (NFW) dark matter halo. If all lenses obey one of these two descriptions, then one would obtain $H_0=74^{+1.7}_{-1.8}\,\hunit$, a 2\% uncertainty (\cite{Millon2020} and references therein). Note that the most precise single lens alone \citep{Shajib2020} yields  $H_0 = 74.2^{+2.7}_{-3.0}\,\hunit$  (4\% precision). 

However, this approach could potentially bias the Hubble constant measurement if the assumed density profile turns out to be incorrect, or underestimate the errors should the profile  deviate significantly from the assumed form on a lens-to-lens basis. Because the distance determination and the inference of the Hubble constant depends so crucially on the lens mass model, attention subsequently turned to relaxing the mass lens modeling assumptions and on treating the mass profile on a lens-by-lens basis.

\cite{Birrer2020} chose a much more flexible parameterization of the radial mass density profile, and, using only the measured time delays and stellar kinematics, finds that the uncertainty on $H_0$ grows from 2\% to 8\% for the TDCOSMO sample, giving $H_0 = 74.5^{+5.6}_{-6.1}\,\hunit$ [3.$b$], but does not change the mean value. Additionally, \cite{Denzel2021} analysed 8 quadruply lensed quasars (with 6 systems shared with TDCOSMO) with a free-form modeling of the lens mass density profile and obtained a slightly lower central value with $H_0=71.8 ^{+3.9}_{-3.3}\,\hunit$. As a further development, \cite{Shajib2023} for the first time break the MSD using spatially resolved kinematics (as opposed to single-aperture kinematics of \cite{Birrer2020}) of a specific lens galaxy (RXJ1131-1231) and implement a mass model that  captures the full flexibility of the MSD. Their findings are consistent with those of \cite{Birrer2020}, with a 9.4\% uncertainty on $H_0$ from a single object. 

\begin{figure}[h]
\includegraphics[width=4.2in, trim = 0.3cm 4.3cm 0cm 4.5cm, clip]{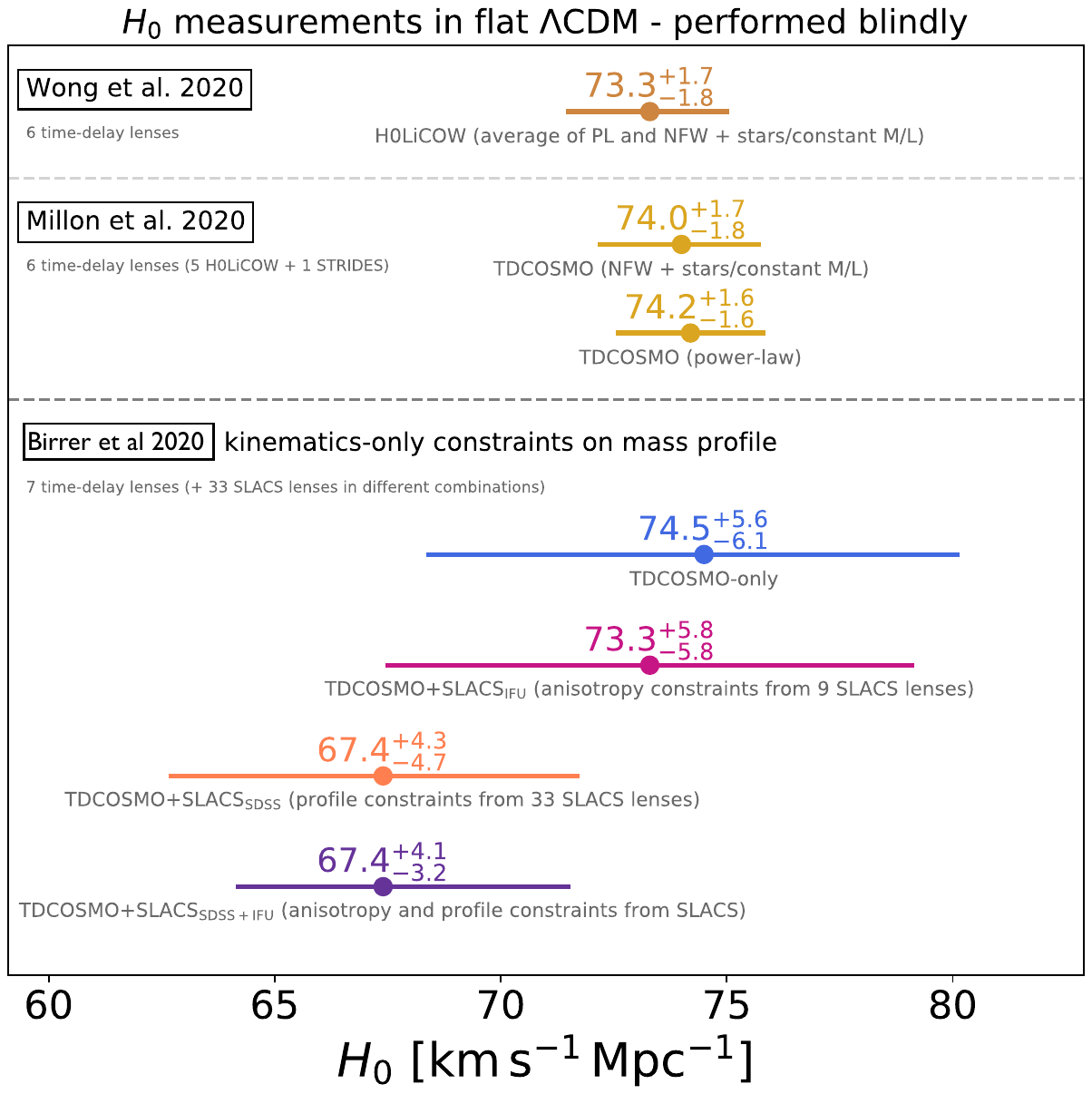}
\caption{Adapted from \cite{Birrer2020}. Summary of $H_0$ constraints from time delay lensing.  The  first three determinations from the top  use a model prior on the deflector mass distribution as detailed in the text. The rest of the determinations rely instead on a data prior. We refer to the main text for more details. }
\label{figTDL}
\end{figure}

Interestingly, \cite{Birrer2020} also introduced a hierarchical framework in which maximally conservative assumptions on the mass density profile are made while external datasets can be combined with the time-delay lenses. This approach is envisioned to become increasingly useful as new large datasets are becoming available (e.g., JWST, LSST, or other complementary  observations and external data sets). They then combined TDCOSMO lenses with stellar kinematics  of a sample of lenses from the Sloan Lens Advanced Camera for Surveys (SLACS), which do not themselves have time-delay information. This shifts the mean recovered value of $H_0$ down  by about $7\,\hunit$ (although still being in broad statistical agreement). When considering this shift one should bear in mind that there are intrinsic differences between the lenses in TDCOSMO and those in SLACS (the two samples differ strongly in redshift-both redshift of the lens and the source, and TDCOSMO is mostly composed of quadruply-lensed systems, while SLACS contains mostly double-lensed systems)  which could lead to systematically selecting different kinds of lenses The SLAC-derived deflector information might not translate to TDCOSMO deflectors, potentially  introducing a bias in the inferred distances. \Cref{figTDL} summarizes  all these findings.

 To recap, time-delay cosmography represents a single-step and direct inference of distances and thus $H_0$. The biggest systematic uncertainty currently arises from the mass sheet degeneracy characterized by the modeling of the mass profile of the lens/deflector.
Adopting a simple mass profile model yields competitive uncertainties for the Hubble constant. With our current knowledge of galaxies, strong lensing time delays agree with high values of $H_0$ to a few percent and 
so far there is no conclusive evidence for a significant deviation of the mass profile from being a simple superposition of an NFW profile with  the observed stellar component or simply a power law.
However, simple models might not be sufficient at this required level of accuracy.

This hypothesis is being verified by allowing more flexible mass profiles to allow full exploration of model degeneracies. Implementing the full MSD in the lens models  is possible, but constraints weaken. Relaxing the mass model at the moment significantly increases the errors on $H_0$ but currently leads to consistent high mean values. 
 The hierarchical modeling approach needs large samples and in this approach,  the time-delay lenses  must match the properties of the other lenses. This is  challenging as there are subtle selection biases to measure each type of lens. At the moment this approach (but with a fundamentally different lensing sample (SLACS)) indicates a possible deviation from high $H_0$ values (but uncertainties are large). A complementary approach is to resort to spatially resolved kinematics. Spatially resolved kinematics of the lens with for example JWST or ELT for a sample of $\sim 40$ lensed quasars or  $\sim 20$ supernovae are expected to achieve $\sim 1\%$ errors on the Hubble constant determination even without making any assumption of the lens mass profiles e.g.,~\cite{BirrerTreu21,Shajib2023}.

\subsection{Standard Sirens} \label{ssec:sirens}
The simultaneous observations of gravitational and electromagnetic signals from compact object mergers at cosmic distances offer a way to bypass the distance ladder \citep{Schutz1986}. This is the essence of the \enquote{standard sirens} approach to measuring the Hubble rate \begin{marginnote}[]
 \entry{Standard siren}{Merging binary emitting detectable gravitation waves signal (and with known redshift).}
\end{marginnote}%
. The gravitational wave signal of a merging binary encodes the luminosity distance to the source which can be calibrated based on a general relativistic model of the inspiral. In fact, for compact binary coalescence, the amplitude of the gravitational wave signal (the strain) is inversely proportional to the luminosity distance. The proportionality factor depends on the gravitational frequency evolution, the combination of the masses of the two objects called \enquote{chirp mass} (which in turn depends on the frequency evolution),  the source position on the sky, and importantly the orientation and polarization. By measuring the amplitude and frequency of the gravitational wave, the luminosity distance can be determined (albeit degenerate with the orientation angle), but there is not enough information to also determine the redshift. For this, an optical (electromagnetic) counterpart is required to pinpoint the host galaxy of the event, the spectrum of which provides a determination of the redshift.

The binary neutron star merger GW170817 \citep{Abbott2017, Abbott2017b} with a kilonova counterpart GRB 170817A which is localized uniquely the host galaxy NGC4993 at $z=0.0097$ provided the first standard siren determination of $H_0=70^{+12}_{-8}\,\hunit$. At this low redshift the luminosity distance directly yields $H_0$ without dependence on other cosmological parameters such as $\Omega_m$\,. For this measurement, the redshift of the host galaxy has been corrected using a simple estimate of the peculiar velocity of $310\pm 150$km/s.

The two dominant contributions to the uncertainty for this initial measurement are the peculiar velocity of the host galaxy and the uncertain inclination of the inspiral event, which is quite degenerate with the overall distance (since both contribute to the overall observed intensity). If one assumes that the emission outflow of GRB 170817A is polar, one can constrain the inclination angle by measuring this outflow. This has been initially done by VLBI radio imaging, which gave an improved constraint for the Hubble constant of $H_0 = 68.9^{+4.7}_{-4.6}\,\hunit$ in \cite{Hotokezaka2019}.
Further incorporating information from the 2M++ large-scale galaxy compilation for the peculiar velocity soon gave a revised value of $H_0 = 68.3^{+4.6}_{-4.5}\,\hunit$ in \cite{Mukherjee2021} (\cite{Nicolaou2020} also demonstrated the large impact of peculiar velocities).

Late observations of the afterglow of the outflow from radio, optical and X-ray lightcurves were used in \cite{Wang2023} to find a higher value of $H_0 = 72.57^{+4.09}_{-4.17}\,\hunit$ [3.$c$] (using the same VLBI of \cite{Hotokezaka2019}, and peculiar velocity of \cite{Mukherjee2021}), and in \cite{Palmese2023} as $H_0 = 75.46^{+5.34}_{-5.39}\,\hunit$ (using different electromagnetic data and different modeling of the peculiar velocity).

While the direct uncertainty on $H_0$ from the standard siren approach might seem high when compared to other determinations, it is important to stress that this determination bypasses any kind of calibration step, and is from a single event only. While the event was a particularly well-suited one (nearby, high signal-to-noise, with electromagnetic counterpart and identification of host galaxy, etc.) other events would not be unexpected, and the statistical uncertainty should decrease roughly inversely proportional to the square root of the observed number of events.

However, the difference in reported mean values highlights the importance of getting systematics under control, in particular from the inclination angle and the peculiar velocity of the host galaxy. If no consensus is reached in this regard, the otherwise so beneficial standard sirens cannot be used to their full potential.

\subsubsection{Dark sirens}
Events without secure optical counterpart are called \enquote{dark sirens}. For these more common events, alternative redshift determinations can be sought, but these are only statistical. For localized sources cross-correlation with galaxy surveys is a promising approach for such a determination of the Hubble constant from dark sirens. For a large number of events, which are not well localized, the perfect mass-redshift degeneracy in the gravitational waveform can be broken statistically by exploiting the form of the binary black hole mass distribution, which is expected to peak around $30-45 M_{\odot}$ (though this will introduce additional systematic uncertainty in the modeling of this distribution).

\subsection{Absolute and relative ages} \label{ssec:ages}

The age of the Universe or the look-back time, as long as homogeneity and isotropy and therefore a FLRW metric is assumed, is  an integral of $H(z)$. Cosmic ages can thus be used to infer constraints on the Hubble parameter or the Hubble constant.

\subsubsection{Cosmic Chronometers}
The Hubble parameter $H(z)$ can be written as a function of the differential time evolution of the universe as a function of redshift:
\begin{equation}
H(z)=-(1+z)^{-1}\left(\frac{dt}{dz}\right)^{-1}
\end{equation}
Since redshift is relatively easy to observe, the challenge is to find a reliable estimator for the differential time $dt$ over a range of redshifts. 
This is the basis of the idea of the {\it cosmic chronometers} method \citep{JL02}. If suitable chronometer objects (homogeneous across cosmic times and robustly and precisely tracing the differential age evolution) across a wide range of redshifts can be observed, then $H(z)$ can be extracted in a model-independent way.

To date, very massive and passively evolving galaxies provide the best cosmic chronometers, as long as their spectra can be obtained over a wide enough wavelength range, with enough resolution, and with enough signal-to-noise. An important underlying assumption is that these can be carefully selected to provide a homogeneous population of cosmological objects with synchronized star formation. In other words, these \enquote{clocks} started \enquote{ticking} at the same time and can therefore be tracers of the differential age evolution of the Universe.

A recent review of the approach and relevant works can be found in \cite{Moresco2022, Moresco23} and references therein. Here we briefly summarize the methodology and report the implications for $H(z)$ and the Hubble tension. Three main ingredients are at the basis of the cosmic chronometer method:
\begin{enumerate}
    \item \textbf{The definition of a sample of chronometers.} The challenge is to find a homogeneous population, and a practical approach is to seek at each redshift  a sample  representing the oldest objects in the universe. There is extensive literature  supporting the case that these are extremely massive and passively evolving galaxies. However, a simple criterion is not able \textit{per se} to select a pure sample of passively evolving galaxies and a combination of several criteria needs to be adopted. 
    \item \textbf{The determination of the differential age $\bm{dt}$.} Different methods have been explored in the literature broadly divided into two philosophies: using the full spectral information or selecting only specific features well localized in wavelength, sensitive to age and relatively insensitive to other properties. All these methods rely on stellar population synthesis models to calibrate the stellar clock. The main uncertainty does not come from the stellar models in itself but rather from the use of libraries of stellar spectra to reproduce the integrated properties of the integrated spectrum that is used to obtain the stellar clock. A detailed study of the impacts of stellar population synthesis models can be found in \cite{MorescoSys}
    
    \item \textbf{Control of systematic errors.} The age determination might be biased by several effects and assumptions. The relevant ones are: small residual degeneracy with metallicity, incorrect prior on the assumed star formation history, assumptions in the stellar population synthesis model, and contamination by a young  stellar component  (also known as frosting).
\end{enumerate} 
High resolution and signal-to-noise spectra are key to greatly reducing or removing degeneracies (age with metallicity, star formation history, or dust content) and addressing these three points.
It is important to note that any systematic effect producing an overall systematic offset in the estimate of the age $t$ is canceled out (or significantly reduced) in the measurement of the differential age $dt$.

 \begin{figure}[h]
\includegraphics[width=4.5in]{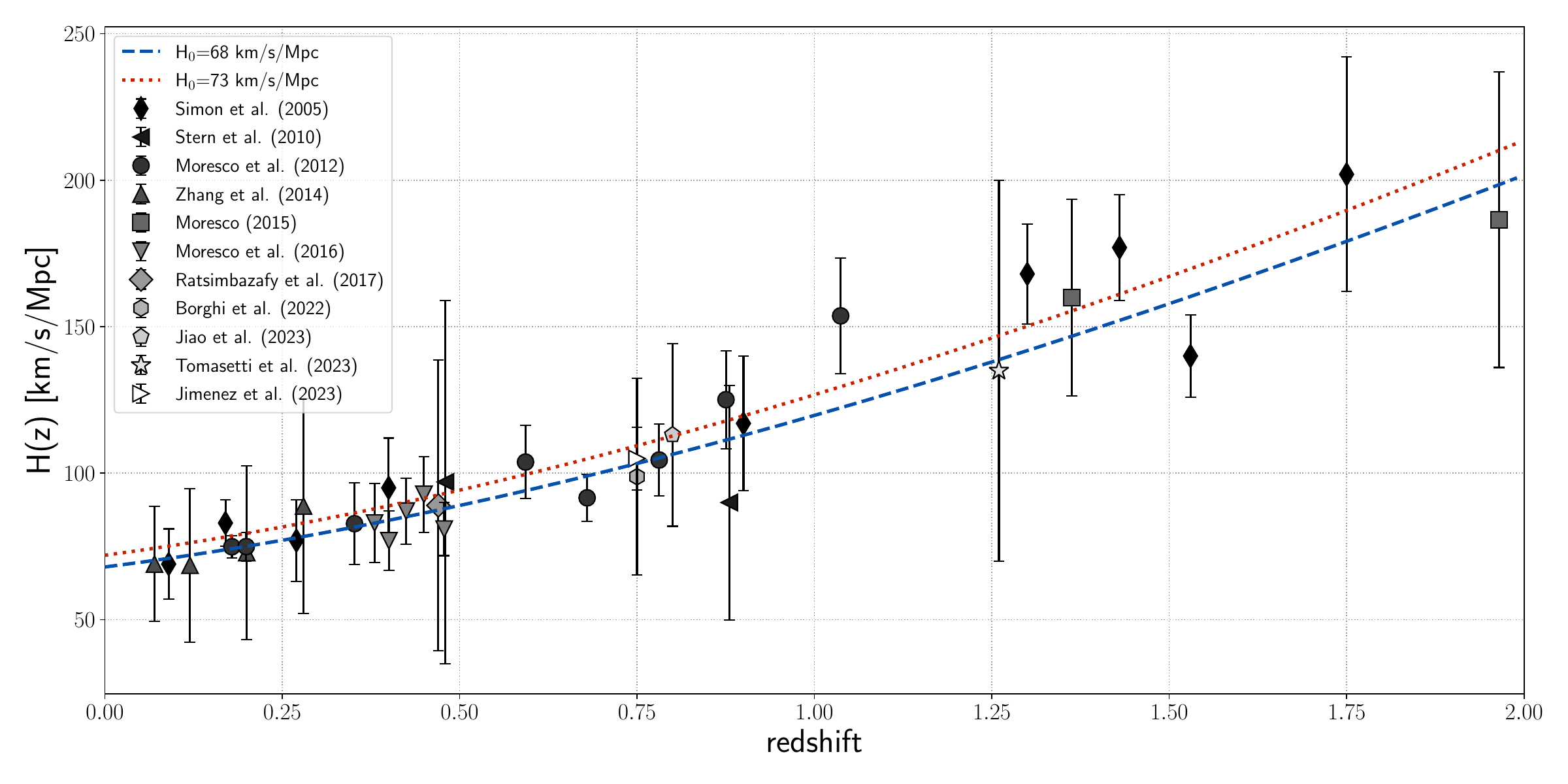}
\caption{Courtesy of M. Moresco, adapted from \cite{Moresco23}. 
Latest compilation of Hubble parameter measurements as a function of redshift obtained with cosmic chronometers. The different symbols correspond to different studies and different samples as indicated in the legend.  The dashed (dotted) line corresponds to the $\Lambda$CDM prediction for $\Omega_m=0.3$ and  $H_0=68 (73)$\, \hunit.}
\label{figHz}
\end{figure}

The state-of-the-art compilation, \cite{MorescoSys}, provides $H(z)$ at 32 redshifts ranging between $0.07<z<1.96$ and their full error covariance, see Fig~\ref{figHz}. From this, the Hubble constant can be obtained by extrapolating $H(z)$ at $z=0$. This has been done within a cosmological model or employing extrapolation techniques which only rely on the smoothness of the $H(z)$ function. In a minimal $\Lambda$CDM model this yields $H_0=66.7 \pm 5.3\,\hunit$ [4.$a$] \citep{Moresco23}.
This $\sim 8$\% uncertainty is dominated by systematics of which the dominant component accounts for the differences among stellar population synthesis models. The size of the error-bars on $H_0$ including systematic uncertainties is still too large to weigh in on the tension. The power of this approach relies upon constraining $H(z)$ across different redshifts in a cosmological model-independent way.

\subsubsection{Building expansion histories}
The differential ages approach is  effectively \enquote{calibrated} (or \enquote{anchored}) at the redshift where the signal-to-noise ratio is highest, which at the moment is $z\sim 0.7$. 
It is also possible to use other measurements of the uncalibrated expansion history $E(z)$ in conjunction with the cosmic chronometers' $H(z)$,
to extract the Hubble constant.
This is commonly done with supernovae, but can also be done with other uncalibrated probes of the expansion history such as quasars or gamma-ray bursts (GRBs). More details on these emerging uncalibrated probes can be found in \cite{Moresco2022}. GRBs use a correlation between their peak energy and the isotropic energy (related to bolometric fluence), which provides an uncalibrated distance measure. For quasars, there are multiple methods, such as the well-known power-law relation between observed UV and X-ray luminosities,\footnote{These objects obey a tight UV/X-ray luminosity relation $\log L_\mathrm{UV} = {\beta} + \gamma \log L_\mathrm{X-ray}$ for which the coefficient $\beta$ is a priori unknown. Given that any measured flux scales with $d^{-2}$, their power-law ratio $\log F_\mathrm{UV} - \gamma \log F_\mathrm{X-ray} = \beta + (2\gamma-2) \log d$ can be measured and used to determine their uncalibrated distance, see for example \cite{Risaliti2019,Moresco2022}.} or the relation between their radius and luminosity.\footnote{These can be probed through mainly two established methods. The first is a time-delay measurement of reverberations from changes of the flux of the central quasar being reflected in spectra of the accretion disk (see e.g., \cite{Khadka2021}). The second is an estimation of the angular size from the visibility modulus (ratio of total flux and fringe amplitude (part of flux that is angularly small compared to the measuring interferometer)), see for example \cite{Cao2017}.}

In combination with SNeIa this approach yields $H_0\sim 68 \pm 3 \,\hunit$ \citep{Haridasu}  with small variations depending on exact reconstruction and extrapolation at $z=0$ adopted. However, the $H_0$ values are strongly dependent on the datasets used for calibration and for the expansion history.\footnote{Results as high as  $H_0 \sim 71.5\pm 2\,\hunit$  can be obtained but using very heterogeneous data sets such as quasar angular sizes using observations dating back to the 1980's. In such cases, it should be stressed that the small quoted uncertainties are primarily statistical, and there is a large undetermined systematic uncertainty that is not examined in the corresponding literature. Also not all measurements use the latest data and full covariance matrix, so care must be taken in interpreting these results.}

\subsubsection{Age of the universe}

The age of the oldest objects in the Universe has long been used to constrain the cosmological model.
The age of the Universe is the lookback time:
\begin{equation}
t_U=9.778{\rm Gyr}/h \int_0^{\infty}\frac{dz'}{(1+z')E(z')} \,,
\end{equation}
In cosmological models that are in rough agreement with the low redshift observations, the integral is dominated by low-redshift contributions ($z \lesssim 10$) where $E(z)$ is reasonably small (it grows as $(1+z)^{3/2}$ during matter domination). In particular, exotic pre-recombination physics does not affect the age of the universe. However, the crux of this probe is that the expansion history in this low-redshift regime is a priori unknown and other cosmological probes have to be used to determine $E(z)$.  Stellar ages  of old individual stars or of coeval stellar populations of the oldest objects in the near universe (globular clusters) have recently provided interesting constraints on the age $t_U$\,. 

Strictly speaking, the age of the oldest objects can only be used as a lower bound for the age of the universe (and thus as an upper bound on the Hubble constant). The underlying assumption is that  the  time gap between formation time and the big bang is small as there is very little time between the big band and $z\gtrsim 10$. 

Using a Bayesian technique to analyze all the relevant features in the color magnitude diagram, \cite{Valcin20,Valcin21} obtained a joint determination of ages, metallicities and distances for 68 globular clusters. This analysis yields\footnote{Adding a small $\Delta t \sim 0.25$ Gyr for formation time between the big bang and the formation redshift at $z\sim 10$.} an estimate for the age of the universe of $t_U=13.5^{+0.16}_{-0.14}{\rm (stat)}\pm 0.23 \mathrm{(sys)}$ Gyr, where the uncertainty  has been divided in statistical and systematic contributions.\footnote{For a basic cross check, 
\cite{Ying2023}, using the full-color magnitude diagram, independently determined the age of M92, one of the  globular clusters used by  \cite{Valcin20,Valcin21} finding good agreement. The $t_U$ determination using the M92 age of \cite{Ying2023} is  $t_U=14.05\pm 0.75$ Gyr.} 

Using only Pantheon+ supernovae to constrain $\Omega_m = 0.334\pm 0.018$ \citep{Brout2022}, one obtains for the age of \cite{Valcin20,Valcin21} in a $\Lambda$CDM model a Hubble parameter of $H_0 = 67.7 \pm 1.2 \mathrm{(stat)} \pm 1.3 \mathrm{(sys)}\,\hunit$ [4.$b$].

It is important to note that this is a relatively recent $H_0$ measurement which, while obtaining competitive error-bars, has not yet undergone the same level of scrutiny as e.g., the determinations from \cref{sec:distance}. A redteaming approach  to $H_0$ for age-based determinations would be  very valuable.

\section{\enquote{WHO ORDERED THAT?}}

\subsection{A matter of anchors}
In summary, the two $H_0$ camps are separated by a  $\sim 7\%$ relative difference, while measurement uncertainties are at the 1\% level (or below). It was qualitatively  clear since  about 2016 \citep{bernal_trouble_2016, soundsdiscordant}  that, within $\Lambda$CDM and popular parametric deviations from this model,  the $H_0$ tension could be also seen as a sound horizon ($r_d$) tension. In all the models considered (which also included variations on the $\Lambda$CDM expansion history), once the model parameters are constrained by a (sub-)set of observations, the model very rigidly connects the properties of the early Universe to the current expansion rate: $E(z)$ is highly constrained (see \cref{sec:expansionhistory}).

It appears to be indeed a problem of anchors:  because the shape of the expansion history is tightly constrained by current data in a model-independent way, the cosmic ladder is quite rigid; it is the early-time anchor $r_d$ (and possibly also the equality scale) calibrated on early-time physics that does
not agree with the late-time \enquote{anchor} of $H_0$.
\subsection{The tension}
\begin{figure}
    \centering
    \includegraphics[width=\textwidth]{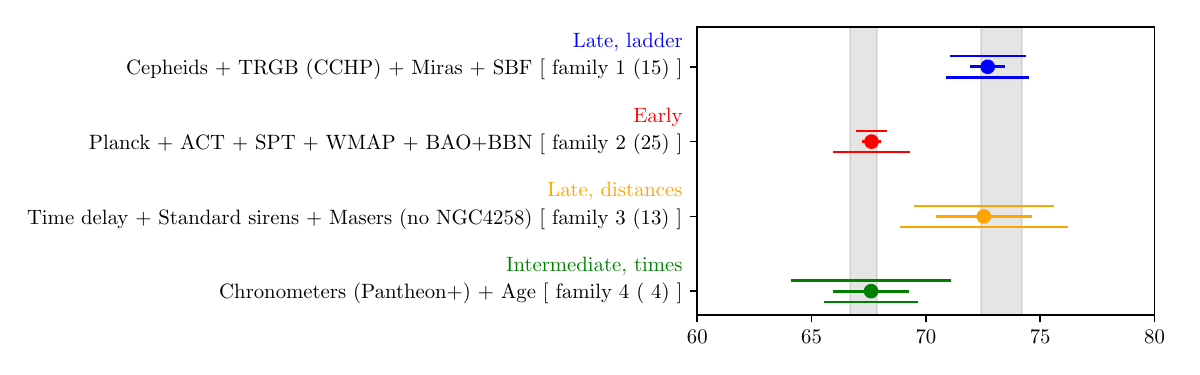}
    \caption{Summary of the $H_0$ values from the four families. For each family we take the weighted mean  and weighted uncertainty of the representative (independent) determinations indicated in the text by the corresponding code [number of the family.letter], as the central value and its error.
    As  estimates of the \enquote{unknown unknown} systematics  we take the \cite{baccus} conservative range of only the independent  representative measurements, (and the scatter of all relevant determinations   {\it \`a la} \cite{Riess2022}, in the legend the number in brackets after the family denotes the number of independent measurements used for this scatter determination) from each family
    and display it as an additional thinner  upper (lower) error bar. To be conservative the CCHP TRGB determination is included in the first family.}
    \label{fig:summary}
\end{figure}

As \cref{fig:summary} summarizes,  the different \enquote{families} of  approaches to $H_0$ yield values that land in  one of the two camps, either low $H_0$ $\sim 68 \hunit$ or high $H_0$ $\sim 73 \hunit$.  We recognize 4 families: 
\begin{enumerate}
    \item The direct distance ladder, denoted \enquote{Late, ladder}, this includes the 15 measurements (of which 4 are the independent and representative ones indicated by [1.{\it a,..,d }]) described in \cref{sec:distance}.  
    \item The inverse distance ladder, also denoted \enquote{Early}, which correspond to the (25 of which 5 representative and independent  indicated by  [2.{\it a,..,e }]) measurements of \cref{sec:invdist}.
    \item Measurements independent of a ladder, such as in \cref{ssec:timedelays,ssec:sirens} which rely on measuring distances.  These are 13 in total of which  3 representative/independent indicated by [3.{\it a,..,c}]. We denote these by \enquote{Late, distances}.
    \item Measurements (4 of which 2 representative [4.{\it {a,b}}]) independent of a ladder, such as in \cref{ssec:ages} which rely on measuring times. We refer to these as \enquote{Intermediate, times}.
\end{enumerate}
We recognize that the megamaser-based $H_0$ determination could belong to \enquote{Late, distances} or to \enquote{Late, ladder}, but as we see below this assignment does not change our conclusions.

The measurements denoted \enquote{Early} are centered in the low (i.e., $68 \hunit$) camp. The measurements denoted \enquote{Late, ladder} and \enquote{Late, distances} are centered in the high (i.e., $73 \hunit$) camp.  The measurement \enquote{Intermediate, times} on the other hand seems to be in the low camp but the high camp due to its large uncertainty is not farther away than $\sim 2.3\sigma$.

Before we move on to discuss the implications of this tension it is interesting to summarize the reactions from the community.

\subsubsection{Reactions from the community} 

The community has been divided since the onset of the \enquote{problem}. On one side the CMB data and the analysis methods have been heavily  scrutinized, but it became quickly clear that   using independent data,  performing independent analyses, and even  completely bypassing the use of CMB data  yield $H_0$ values consistently in the 68 $\hunit$ camp.
To change the $H_0$ value  inferred from the early Universe it is the model that needs to be changed. 
The spectacular success of the  standard cosmological model over the last two decades, with the exquisite  consistency between CMB and large-scale structure data -- mapping the majority of  the Universe's history -- within that model makes this hard to accept.  
Therefore  part of the community initially chose the position which we (tongue in cheek) denote \enquote{this is fine}. The reasoning behind this is that the  measurements are challenging  and the model works so well  that the measurement uncertainties (especially those of the local $H_0$  determinations) should be increased (see e.g., \cite{baccus}), a position in which some philosophers of science seems to share (see e.g., \cite{Gueguen}).  
However, as the tension reaches and surpasses the 5$\sigma$ threshold, this position becomes increasingly untenable.
Another part of the community (mostly the theoretical sub-community) has embraced the emerging tension and vigorously and extensively  explored theoretical solutions, all representing deviations from the standard $\Lambda$CDM model and therefore new physics.  After all, we have to recognize that the $\Lambda$CDM model, despite its  deep connections to fundamental physics and  all its successes, is  ultimately phenomenological. It establishes a robust framework, in which, however, fundamental issues remain unresolved. Dark energy and dark matter are {\it ad hoc} components. It is not unreasonable to think that (much like the epicycles or the aether in the past) the model they feature in might be an effective model and these components place-holders for something else deeper. This has been the driver of these theoretical efforts.
 We point to the excellent reviews \cite{DiValentino:2021izs,H0olympics} for an overview of the theoretical solutions.
A  recent  overview of the community positions on the tension is presented also in (Riess et al in prep.)\cite{Riess23} where the author suggests proceeding with \enquote{curiosity}.  

In this vein, to guide this curiosity, it is useful to review what guardrails are now in place, set by  the extensive body of literature on the matter produced over the past decade. In the quest for a solution to the tension guardrails do not tell us if we are on the right road, but prevent us  from going off-road completely.

\subsection{Guardrails and lessons learned}
\label{ssec:guardrails}
It is clear that there are two competing  explanations for this tension: either the  interpretation of the distance measurements is incorrect or the cosmological model we use to model early-time physics and  interpret the observations (CMB or BAO+BBN or both)  to infer a value for $H_0$ is incorrect. 
It is now  also clear  that  to change the sound horizon by the 7\% required to solve the tension, there is limited room for manoeuvre: it requires introducing  some  new  physics in the e-fold of expansion just before recombination (see \cite{Knox_Millea20}). However, this comes at a cost. All attempts so far spoil the good agreement with observations that the $\Lambda$CDM model enjoys (see also \cref{ssec:graveyard}). 

Finding a solution to the Hubble tension that does not rely on measurements being incorrect requires finding a cosmological model that can fit all available cosmological observations at least as well as the $\Lambda$CDM model does.  This also in turn requires not making  other low-level inconsistencies worse, as we illustrate below.

\subsubsection{Implications for other tensions}

The Hubble tension should not be viewed entirely on its own, as constraints for other cosmological parameters from the early Universe are dependent on the assumed underlying model. Perhaps the most straightforward example is the matter fraction $\Omega_m$\,, which is strongly correlated with the Hubble constant in CMB anisotropy analyses due to the tight constraint on $\Omega_m h^3$ from the sound horizon angle, as discussed in \cref{ssec:CMBmain} and supplemantal material \cref{ssec:CMB_SM}. As such, theory-based solutions to the Hubble tension typically cause also changes in the matter abundance, such as for example when a variation of fundamental constants is employed in order to shift the time of recombination (which reduces the sound horizon, allowing to raise $H_0$ to keep the sound horizon angular extent the same), see \cite{Lee2023}. The typical problem of changing $\Omega_m$ is that both supernovae and  uncalibrated BAO already give precise measurements of $\Omega_m$ in good agreement with the usual $\Lambda$CDM value.

A similar (albeit a bit more complicated) problem is encountered for the $S_8\sim \sigma_8\sqrt{\frac{\Omega_m}{0.3}}$ clustering parameter well determined from weak cosmological lensing (for the latest constraints see e.g., \cite{KidsDES2023}).  The $\sigma_8$ parameter quantifies the standard deviation of matter overdensities when smoothed on scales of $8$ Mpc/$h$. Weak-lensing observations (such as e.g., \cite{KidsDES2023}) tend to report a lower value of $S_8$ compared to that estimated from CMB data \citep{PlanckParams2020}. For a recent review see \cite{DiValentinoS8}.
The $S_8$ predicted by a model that is calibrated on the observed CMB anisotropies is impacted by a change of $H_0$  for two reasons: {\it i)} because the Hubble constant rescales physical densities (such as the dark energy density causing the decay of gravitational potentials towards $z\to 0$, which impacts how well dark matter can cluster and perturbations can grow) and {\it ii)} because a change in $H_0$ modifies the matter radiation equality (via the $\Omega_m-h $ degeneracy), thus the shape of the matter transfer function.\footnote{The redshift of matter radiation equality is proportional to $\Omega_m h^2$, but since only $\Omega_m h^3$ is approximately fixed by the CMB angular sound horizon, a variation of $h$ also shifts $\Omega_m h^2$.}
Even beyond these effects, a variation of $h$ also shifts the smoothing scale which defines the $\sigma_8$ parameter, since this scale is given by $8 \mathrm{Mpc}/h$. Overall, this results in a dependence of $S_8 \propto h^{-2.25}$ (when fitting the CMB data within a $\Lambda$CDM model). As such, the mild tension in $S_8$ would be eased in the $\Lambda$CDM model if the fit to CMB data were to return a higher $H_0$ around $\sim 70.5\,\hunit$ (though this is rather unlikely, see \cref{ssec:cmb_systematics}). Problematically, most early-time solutions to the Hubble tension, once calibrated on CMB data cause only a small downward or even upward shift of the predicted $S_8$\,. As such, the two tensions are linked quite tightly, and proposed (early time) deviations from $\Lambda$CDM  to  ease  the Hubble tension immediately impact the $S_8$ tension.\footnote{Conversely, since many solutions to the $S_8$ tension typically act on the small scale power spectrum, they do not necessarily significantly impact the  $H_0$ determination from the CMB.}

\subsubsection{The graveyard and lessons learned}\label{ssec:graveyard}

A large variety of models have been proposed to ease the Hubble tension (a review can be found in \cite{DiValentino:2021izs}). While many models initially seemed rather successful, it has become clear that naive modifications are not allowed once  cosmological data from different epochs is used. A good example of this is the model of a curved universe,\footnote{A similar conclusion can be made for a variation of the dark energy equation of state. Here one finds $w_0 = -1.58^{+0.52}_{-0.41}$ from primary CMB alone, and $w_0 = -1.04\pm 0.10$ when BAO and CMB lensing are added.} which  exploits the  large  geometric degeneracy  in the CMB (see \cref{fig:H0_whisker_cmb}). However, the curvature required for significant changes of $H_0$ is rather inconsistent with the tight constraints enabled by BAO or SNeIa ($\Omega_k = -0.044\pm 0.033$ from primary CMB, and $\Omega_k = 0.0007\pm 0.0037$ once BAO and CMB lensing are added, see \cite{PlanckParams2020}). As such, it is important to consider the \enquote{graveyard} of unsuccessful models in order to pin down which model-building solutions to the Hubble tension remain viable.
We hope that this small survey of proposed (and interesting) solutions that in fact don't work is useful  to demonstrate that  sometimes, in physics, things that are wrong or do not work can be, in fact, useful. 

Since the angular size of the sound horizon is measured to very high precision from the CMB (see \cref{ssec:CMBmain}), it has to remain almost fixed for any solution attempting to reconcile the Hubble tension. This allows separating the solutions into two categories: early-time solutions with modifications around and before recombination ($z \gtrsim 1100$) and late-time solutions with modifications at a later time (see also \cite{H0olympics} for a comparison of these two types of solutions). The problem of increasing the Hubble constant within a model can then be posed as \enquote{threading a needle from the other side of the Universe}. In this analogy, the thread begins at a fixed location (the sound horizon of the early Universe) and through its subsequent path (angular diameter distance) we have to reach the needle's eye (a given value of $H_0$).

The late-time solutions by definition leave the sound horizon physics the same, and only modify the angular diameter distance under which we observe the standard ruler. This is the essential problem of these late-time solutions, they have to carefully balance the increase of the Hubble parameter with a decrease of the expansion rate to keep the angular diameter distance fixed to the value preferred by the CMB. In other words, they have to carefully perturb the thread to fit a new needle eye at a higher $H_0$ value.

Importantly, these late-time solutions are constrained by the wealth of late-time observations, such as the BAO or the SNeIa, which tightly constrain the expansion rate $E(z)$ in the intermediate regime and thus the path the thread can take (see also \cref{sec:expansionhistory}). This is particularly difficult since the expansion rate has the largest impact on the angular diameter distance only at the smallest redshifts $z \ll 10$, where these measurements are most abundant. Late-time solutions thus have to thread the needle's eye with a large part of their thread being stiff.
 Several attempts have also  been made  to modify the expansion rate at $z \lesssim 0.1$ in order to break the assumptions underlying the distance ladder. However, the SNeIa and other low redshift probes by now paint a similar picture that deviations from a smooth expansion history are extremely unlikely, and such hockey-stick expansion rate models are largely excluded \citep{Camarena2020}.

Similar considerations apply to the attempt to postulate that we observe the Universe from a giant void. Such an underdense region would in principle lead to a faster local expansion/recession rate than what the Hubble flow would predict, explaining why late time local $H_0$ values are higher than the \enquote{global}  one. Only a \enquote{designer} void profile could avoid a hockey-stick expansion rate. Moreover the resulting pattern of peculiar velocity (voids expand) would produce other signatures and is overall heavily disfavored \citep{kSZ2011,Camarenavoid}.
We also stress that this type of solution involves dropping the Cosmological principle and the Copernican principle. Not only we, the observer, would have to be placed in a special position, but also the underdensity needed to achieve the required shift in $H_0$ would have to be quite extreme. Also the assumption of statistical homogeneity  would have to be dropped, which would mean dropping the FRLW metric too.

The early-time solutions have a somewhat simpler goal of changing both the needle's eye and the starting location. If this is done coherently, no modifications to the path of the thread need to be made, leaving intermediate probes such as BAO or SNeIa unchanged. As such, a big part of the model building focus has shifted to early universe solutions which modify the sound horizon and the Hubble parameter, but otherwise leave the expansion rate (and thus angular diameter distance) at low redshift invariant. The difficulty here is not disturbing the excellently measured baryonic acoustic oscillations of the CMB.
Problems with this approach typically arise at the level of the perturbations, for example in an over-damped tail of the CMB from additional dark relics or from varying fundamental constants.
 Different pipelines for extracting the angular power spectrum from the underlying CMB maps can have an impact of how strongly these perturbations are constrained, such as in the case of early dark energy \citep{231019899, 231100524}.

Modifications of the interpretation of the sound horizon standard ruler as seen in the BAO have also been attempted, for example postulating a large velocity bias between dark matter and baryons  \citep{BAOvelocitybias1}. Subsequent observations have placed stringent limit on the maximum magnitude for  this effect showing that it is too small to matter~\citep{BAOvelocitybias2}.

This has to be a critical lesson for anyone attempting to propose a model-based solution for the Hubble tension: if the solution is to be widely accepted, it needs to conform to the measurements of {\bf all} available observations including especially the full CMB temperature and polarization data and  the expansion history as constrained  from BAO and SNeIa.

\section{CONCLUSIONS}
The local expansion rate of the Universe, measured in the low redshift Universe does not agree with its  value as inferred from early Universe observations when interpreted within the standard cosmological model. 

This discrepancy, denoted \enquote{Hubble tension}, is one of the most intriguing problems in cosmology today, becoming the \enquote{make-or-break} test for the $\Lambda$CDM model: a high-stakes situation with potential implications well beyond cosmology.

There is now a considerable number of qualitatively different $H_0$ determinations: some directly measure the relation between cosmological redshifts and distances, others constrain a global parameter of the cosmological model, and others how time passes with redshift.  Each measurement has a different meaning, relies on different physics and  assumptions, and should be viewed as a different  quantity. We have argued that, in fact, there are four qualitatively different families of measurements, each relying on a different, internally coherent set of assumptions (although some key assumptions are in common to all). Crucially each family covers very different data and analyses methods, although, admittedly,  the level of scrutiny of the different $H_0$ determinations is  certainly not the same across families. The variation within families gives a qualitative estimate of observational/analysis systematics, while the variation across families offers a route to test the model’s assumptions.
The  measurements from these different families are expected to  agree only if there are no  significant observational/analysis systematics and if all assumptions (including the adopted cosmological model) are (sufficiently) correct.
We have shown that, at high statistical significance, these families of measurements cluster around two well-separated \enquote{camps}. 
There is a pattern, however (\cref{fig:summary}).
 
Each family has a series of internal consistency checks built into it. These could certainly be improved and made more robust with more data, of course. A red-teaming approach will certainly  strengthen the robustness of each approach.  Reproducibility can be further fostered  by improving public availability and accessibility  of data and software (and detailed documentation), facilitating independent confirmation by independent teams.  Such effort is not for free, is time expensive and labour intensive and not always favorably recognized, but it is crucially important. Some research areas have embraced this more than other.

Model-independent (or model-agnostic) approaches to constrain $H_0$ can be combined with model-dependent ones to test specific model ingredients. Blind analyses, although not a silver bullet, protect against experimenter's bias (which is statistically weighted towards the standard cosmological model, but not always). 

The cross-family distance has reached the $5\sigma$ statistical  level, which cannot be ignored.
Because of the internal consistency checks and diversity of data/and methodology within each family,  an explanation that involves  data-related systematics starts becoming unlikely (it would require several unrelated  systematic effects to have been all underestimated  and to all combine in the same direction).

On the other hand it is not unreasonable to accept that  the $\Lambda$CDM model, as successful as it has been, is an effective model, with several {\it ad-hoc} components. It is based on a whole suite of assumptions and is extremely predictive, and so -- if sufficiently stressed -- might show hairline fractures if not cracks.
It is therefore instructive to think of the Hubble tension in this light and consider  also other possible hairline fractures, low level tensions  of the model that  in isolation are not structural faults but may be all related.

Theoretical solutions involve \enquote{new physics} which, in order not to exacerbate other low-level tensions and preserve the good agreement of $\Lambda$CDM with all other cosmological observations,  seems to be limited to two directions: either modify the early Universe physics, or modify the model at a more basic level, possibly the pillars on which the model is built. None of these options is easy to accept.

On-going and forthcoming observational effort  (Euclid, DESI, JWST, CMB-S4, etc.) all promise to significantly improve $H_0$ determinations for all families.  This will either restore $\Lambda$CDM or  provide  clear guardrails to guide us in the quest for a solution.

\section*{DISCLOSURE STATEMENT}
The authors are not aware of any affiliations, memberships, funding, or financial holdings that
might be perceived as affecting the objectivity of this review. 

\section*{ACKNOWLEDGMENTS}

LV thanks  Dillon Brout, Fred Courbin, Sherry Suyu, Adam Riess and  Stefano Casertano for stimulating discussions and feedback on specific sections of the manuscript at the draft stage. This work was greatly improved thanks to your thoughtful and constructive feedback.

We acknowledge support of \enquote{Center of Excellence Maria de Maeztu 2020-2023} award to the ICCUB (CEX2019-000918-M funded by MCIN/AEI/10.13039/501100011033).
HGM acknowledges support through the program Ramón y Cajal (RYC-2021-034104) of the Spanish Ministry of Science and Innovation.


%


\bibliographystyle{ar-style2}

\include{ads_macros.tex}

\section{Supplemental material}
\subsection{Other approaches to the distance ladder}\label{ssec:otherladders}
\subsubsection{HII galaxies}\label{ssec:hii}
These galaxies show strong Balmer (HII) emission lines, whose intrinsic luminosity strongly correlates with the line velocity broadening. This is the broadening of any given thin emission line caused by a variety of objects with slightly different peculiar velocities (similar to thermal broadening but caused by  the peculiar velocities of objects, not the thermal velocities of atoms). The reason for this correlation of velocity and line-broadening is that in young massive clusters undergoing bursts of star formation the number of ionizing photons is correlated to the corresponding turbulent velocity dispersion of the ionized gas. The same phenomenon has also been observed in extra-galactic HII regions. One major drawback of this method is the need to determine the intrinsic thermal broadening (which has to be subtracted from the overall broadening along with instrumental broadening and fine-structure broadening), for which a good estimate of the gas temperature is required. Another possible drawback is the non-negligible scatter in the relation that can only partially be explained by corrections from radius, color, or metallicity \citep{Chavez2014}. Nevertheless, the method allows for another alternative distance calibration up to relatively large distances, which can be used to measure the Hubble constant. 

In \cite{Fernandez2018} the luminosities are calibrated from the measured flux and known distance to 13 nearby galaxies (hosting 36 objects, with uncertainty dominated by Cepheid distances), and they find $71.0\pm2.8 {\rm (stat.)}\pm2.1 {\rm (sys.)}\,\hunit$. 

\subsubsection{Baryonic Tully Fisher relation}\label{ssec:btfr}
The baryonic Tully Fisher relation (bTFR) can be used to relate measurable rotational velocities of an object with the total baryonic mass. Compared to the classical Tully Fisher relation (involving the total stellar mass) the scatter is much smaller, though still various corrections need to be applied. The mass, on the other hand, is tightly correlated with the absolute luminosity of the galaxy. \cite{Kourkchi2020} uses distance calibrations from Cepheids and the TRGB, deriving a value of $H_0 = 76.0 \pm  1.1\mathrm{(stat)} \pm 2.3\mathrm{(sys)}\,\hunit$. Using instead distances from CosmicFlows-3 \citep{Tully2016} the authors of \cite{Schombert2020} find  $H_0=75.1 \pm 2.3\mathrm{(stat)} \pm 1.5\mathrm{(sys)}\,\hunit$, though it should be mentioned that the CosmicFlows collaboration typically establishes their distances also on the basis of Cepheid/TRGB calibration and the bTFR (with a fixed relation between gas mass and luminosity \cite[Eq.~(3)]{Kourkchi2022}, and for the stellar mass using a given mass-to-light ratio derived using $H_0=75\,\hunit$ distances).\footnote{Given that the focus of CosmicFlows is \enquote{the study of deviations from cosmic expansion} \citep{Tully2023}, this is not an issue for CosmicFlows, though this has to be understood when using their distances in another context.} As such, the Hubble measurements should not be seen as entirely independent from the Cepheid or TRGB calibrations and there is a dependence on assumptions about how the respective baryonic masses relate to the intrinsic luminosity. 

\subsubsection{Fast radio bursts}\label{ssec:frb}
The study of fast radio bursts has recently emerged as another way to indirectly infer the distances to their host galaxies. Fast radio bursts (FRBs) are transient emissions of extremely bright radio pulses that typically last milliseconds, and are believed to originate from magnetars \citep{Bochenek2020}. The emitted radio pulses typically experience dispersion (frequency-dependent delay time) when encountering gas and the corresponding dispersion measure can be well measured and provides a tracer of the column density of the intervening gas distribution. This measure of column density can in principle be combined with assumptions about the intervening gas (such as the fraction of baryons in the IGM) to allow for an inference of the distance, which (once calibrated) can be used to infer the Hubble parameter. Additionally, assumptions about the dispersion measure of the host galaxy and the Milky Way have to be made.

Recent analyses with less restrictive assumptions and broader priors obtain large uncertainties, such as  $H_0 = 73^{+12}_{-8}\hunit$ \citep{FRBJamesetal2022}  or  $H_0 = 62.3 \pm 9.1\,\hunit$ \citep{Hagstotz2022}.\footnote{Newer analyses like \cite{Wu2022,Gao2023} obtain smaller uncertainties around $\sim 6\,\hunit$ by combining with additional information (from simulations or supernovae), but are not yet decisive in terms of the Hubble tension.}

\subsection{Primordial oscillations: simple linear physics yielding an anchor} \label{ssec:primordial_oscillations}
After BBN, a hot plasma of atomic nuclei, free electrons, and vastly more abundant photons constituted  a tightly coupled photon-baryon fluid that filled the Universe. The other species in the $\Lambda$CDM model were three neutrino types (non-interacting and relativistic) and collisionless cold dark matter, interacting only through gravity. Any contribution of dark energy (or cosmological constant) was entirely negligible. Hereafter we describe the situation as in the $\Lambda$CDM model, but stress that modifications to the physics of the acoustic oscillations themselves are tightly constrained by the CMB and BAO probes: at present, there is no evidence of any significant variation from this primordial composition,  in the standard cosmological model this list is complete, and any other component would represent \enquote{new physics}.

The density of the primordial Universe was exquisitely uniform, with perturbations below the $10^{-5}$ level. These perturbations were adiabatic\footnote{The fractional density perturbation is proportionally  the same for all components.} distributed according to a Gaussian random field, with a primordial power spectrum, as seeded by inflation, $P_\mathrm{prim.}(k) = A_s (k/k_p)^{n_s-1}$ with some amplitude $A_s$ and tilt $n_s$ and some arbitrary pivot wavenumber commonly set to $k_p = 0.05/\mathrm{Mpc}$.
 At present there is no strong indication of a deviation from these properties of the primordial perturbations;\footnote{Some large-scale anomalies are present in the CMB anisotropies  at low statistical significance, a topic we revisit in \cref{ssec:CMBmain}.} these properties, although empirically confirmed,  are built-in as part of the standard cosmological model. 

During the Universe evolution (even only the first 400\,000 years) the power spectrum of density perturbations  gets modified from a pure power law. In particular, during radiation domination, $z\gtrsim 3300$, (dark matter) density perturbations grow gravitationally  only outside the horizon:  inside the horizon perturbation growth is strongly  suppressed. The expansion of the Universe and the evolving  size of the horizon imprint a characteristic shape on the matter power spectrum. During matter-domination  dark matter perturbations  can grow  coherently at all scales and the broad-band shape of the matter power spectrum is thus frozen (until perturbations become non-linear at $z<$ a few, see below). This causes the scale that enters the Hubble horizon during the matter-radiation equality, (equality scale), to act as a possible standard ruler scale see \cref{ssec:equality_scale}.

We stress that the simple analytical arguments presented in this chapter are sufficient to gain understanding of the  physics at play. For quantitative predictions, precise enough for data analysis, the full combination of Boltzmann and Einstein equations need to be solved as done by the Einstein-Boltzmann solving codes such as \textsc{camb} \citep{Lewis:1999bs} or \textsc{class} \citep{class1}.

To understand the effect of the presence of adiabatic  perturbations  on a tightly coupled photon-baryon fluid in the early Universe, it is useful to think about them schematically as a spatial superposition of Dirac delta functions  and then focus on a single one, imagining it isolated in an otherwise homogeneous background. The neutrinos  basically free-stream away. But the pressure of the baryon-photon fluid drives a sound wave that expands at sound speed $c_s$, which is close to $c/\sqrt{3}$ with $c$ being the speed of light. The dark matter perturbation stays put.
As the Universe's expansion continued, the temperature of the  plasma continued to drop until photons lacked on average the energy to photo-ionize newly formed nuclei. Recombination (the formation of neutral atoms) halted the scattering of the photons (decoupling)  and allowed them to freely travel, forming the last scattering surface which we observe as the CMB.

Once the photons and baryons decouple, it means that the acoustic waves in the photon-baryon fluid stop propagating, having traveled by a distance equal to the integral of velocity over (conformal) time, a quantity commonly denoted as the sound horizon
\begin{equation}\label{eq:sound_horizon}
    r_s = \int_{z_\mathrm{rec}}^\infty \frac{c_s(z')\mathrm{d}z'}{H(z')}
\end{equation}
where $z_\mathrm{rec} \sim 1100$ is the redshift of recombination.\footnote{We used the relation between time and redshift established by the Hubble parameter, $H = \mathrm{d}t/[(1+z)\mathrm{d}z]$. This relation is simply a result of the definition of the Hubble parameter and redshift with $1+z = 1/a$ and $H = \mathrm{d}\ln a/\mathrm{d}t$, where $a$ is the scale factor.} 

The combined effect of the superposition of all the primordial perturbations implies that  oscillations of the primordial plasma were then imprinted in small temperature anisotropies of the CMB. These are the CMB acoustic peaks, clearly visible in the CMB angular power spectrum. Indeed, the measurement of the CMB angular power spectrum directly gives an angular measure of the sound horizon, as discussed in \cref{ssec:CMBmain}. Moreover, these oscillations are also imprinted in the total matter density, given that the baryons (which at decoupling are displaced from the dark matter overdensity) constitute about 1/5 of the total matter density (i.e., not only baryons fall into the dark matter potential wells but also the dark matter falls into the baryons potential wells).  These are the Baryon Acoustic Oscillations (BAO) visible in the galaxy (and other tracers) power spectrum offering an excellent independent confirmation of the primordial oscillation physics. 
While the photons stop their baryonic acoustic oscillations already at recombination, the baryons continue to feel the effect of the photons until the drag epoch.\footnote{Recombination occurs when the photon optical depth (number of average collisions per photon) reaches unity, while baryon drag occurs when the baryon optical depth reaches unity. 
Since there are many more photons than baryons, the baryons continue to scatter for slightly longer, until around $z_d \sim 1050$.} Since recombination and baryon drag are typically tightly related and close in redshift ($\Delta z \lesssim 50$), in the literature they are sometimes not distinguished. The cosmic standard ruler is thus the sound horizon at recombination $r_s$ for the CMB and the sound horizon at radiation drag (which we denote as $r_d$) for the BAO, and they are tightly related.

The precise way that the oscillations occurred and the way we observe them today (that is, the observed length of the standard ruler) is intrinsically linked to the underlying parameters of the cosmological model. 
The sound horizon itself depends on the photon-to-baryon ratio (more heavy baryons, lower sound speed), time of decoupling (or radiation drag), and the expansion rate of the Universe from the big bang to recombination/drag which in turn depends on the composition (dominated by the radiation and matter components).

As mentioned above, the size of the Hubble horizon at matter-radiation equality could, in principle, also be used as a standard ruler. The matter power spectrum shows a characteristic bend or \enquote{knee} whose position is determined by the epoch of matter-radiation equality, also called the equality scale. The detailed shape of the knee however depends on the details of the matter radiation equality, expansion history of the Universe at that epoch, detailed neutrino properties, and the influence of baryons (also called \enquote{baryonic suppression}). The knee is also not a sharp feature and therefore at the present time, using it as a standard ruler is challenging. However, the shape of the knee can be used  as an anchor of an inverse distance ladder. We return to this point in \cref{ssec:equality_scale}.

It is important to note that the physics of the early Universe (from the first few minutes to recombination) as explained in this section is remarkably simple, can be treated in linear perturbation theory, is extremely well understood, and convincingly observationally verified. The Universe's composition was simple, and the interactions well known. Even dark energy was unimportant. Observations of the early Universe have provided a spectacular confirmation of all the ingredients of the standard cosmological model that impact these early Universe probes.

\subsection{CMB anistropies}\label{ssec:CMB_SM}

To understand the role of CMB observations in yielding a measurement of $H_0$\,, consider that 
the angular power spectrum of temperature and polarization anisotropies, with the rich phenomenology enclosed in the CMB acoustic peaks, has been measured with spectacularly increasing precision and accuracy over the past couple of decades. 
The WMAP mission (2001-2012, \cite{WMAP9yr}) has measured the full CMB sky in 5 frequencies over 9 years producing maps of the CMB anisotropies with $\delta\theta \lesssim 0.3^\circ$ angular resolution, which implies a maximum multipole for the CMB power spectrum of $\ell\simeq1000$ ($\ell \simeq \pi/\delta \theta$). Finer scales were mapped by partial-sky ground-based experiments -- of particular relevance are the Atacama Cosmology Telescope (ACT) and the South Pole Telescope (SPT). While ACT covered $> 15\,000 $deg$^2$ and pushed the angular resolution to $\ell \sim  4000 $ \citep{ACTDR4}, SPT reached $\sim 2000$ deg$^2$ and $ \ell \sim  3000 $ \citep{SPT2022}. Finally the satellite-based Planck surveyor (2009-2013, \cite{Planck2018_overview})  mapped the full CMB sky in 9 frequencies over about 4 years.  Its high angular resolution meant that the CMB anisotropy power spectrum could be measured up to $\ell\sim 2500$. 

While primordial anisotropies dominate the CMB sky at $\ell < 1000$ (these are often referred to as \enquote{primary} anisotropies),\footnote{On these scales the processes of deflection and rescattering of photons from intervening matter after recombination modify the appearance of primary ainistoropies:  reionization, which likely happened at $z\gtrsim 10$,  suppresses coherently the anisotropies at $\ell \gtrsim 40$; the Integrated Sachs-Wolf effect from changes to the photon energy due to gravitational potentials, is relevant at $\ell \lesssim 30$. These are accounted for from theoretical calculations and are often considered an inherent part of the CMB since they cannot be separated by a different frequency dependence.}  at smaller scales physical processes taking place  post-recombination start dominating making it very difficult to access the primordial information. The most relevant one, which cannot be separated out via its frequency dependence, is  weak gravitational lensing, (affecting $\ell \gtrsim 1000$).  The gravitational lensing  signal can be separately constrained by using properties of the primordial perturbations, leading to a measurement of the gravitational lensing potential that can be used in combination with the \enquote{primary} CMB anisotropies.

Galactic and extragalactic foreground emission can contaminate the CMB light,  but for this review it will suffice to say that these are separated out during analysis and interpretation through their frequency dependence.  

\subsubsection{Physical interpretation}\label{ssec:CMB_physintyerp}
The acoustic peaks series appear to subtend a characteristic angle on the CMB sky. We can express this angle as a ratio of the lengths
\begin{equation}
    \theta_s = \frac{r_s}{r_A}~.
\end{equation}
Both the sound horizon $r_s$ and the comoving angular diameter distance $r_A$ in principle encode information about the Hubble constant. The angular diameter distance to the last scattering surface can be written as
\begin{equation}\label{eq:angular_diameter}
    r_A = \frac{(1+z_\mathrm{rec})}{H_0} \int_0^{z_\mathrm{rec}} \frac{dz'}{E(z')}
\end{equation}
which shows that it is inversely proportional to $H_0$ for a given expansion rate (since $H_0$ normalizes the overall Hubble rate). The expansion rate is then typically related to the fractional components of the Universe, such as $\Omega_m$ in the $\Lambda$CDM model. Within that model, one finds $r_A \propto H_0^{-1} \Omega_m^{-0.4}$ to percent level precision. Instead, the comoving size of the sound horizon, eq.~\cref{eq:sound_horizon}, depends strongly on the Hubble rate before recombination. The early time  $H(z)$ is influenced in the $\Lambda$CDM model mostly by the early time physical radiation density $\Omega_r h^2 \propto T_\mathrm{cmb}^4$, but also the physical matter density $\Omega_m h^2$ (the time at which matter begins to dominate the expansion rate $z \sim 3300$ is shortly before recombination $z \sim 1100$). The sound speed, on the other hand, is impacted by the physical baryon abundance $\Omega_b h^2$ (which is also well determined from light element abundance measurements from BBN) and physical photon abundance.\footnote{In principle there is also a dependence on the time of recombination, which itself depends on the cosmological parameters. This dependence is comparatively small, though.} Together the integral for the sound horizon conspires in $\Lambda$CDM to give a dependence of roughly $(\Omega_m h^2)^{-0.25} (\Omega_b h^2)^{-0.13}$. This means that overall the angular sound horizon depends in $\Lambda$CDM roughly as $\theta_s \propto \Omega_m^{0.15} h^{0.5} (\Omega_bh^2)^{-0.13}$. This illustrates how,  even in  $\Lambda$CDM,  there are many degeneracies preventing  directly inferring $h$ from the sound horizon angle. 

Fortunately, the CMB is subject to various additional effects that depend on the Hubble constant, 
such as the height of the first peak that is intimately tied to the redshift of matter-radiation equality ($z_\mathrm{eq} \propto \Omega_m h^2$),\footnote{The height of the first peak is influenced by the early integrated Sachs Wolfe effect, which is the decay of gravitational potentials as the universe transitions from radiation domination to matter domination. For a smaller redshift of matter radiation equality, the potentials at recombination are less decayed (there is a larger potential envelope).} the ratio of odd and even peaks (tied to the baryon-to-photon ratio $R \propto \Omega_b h^2$), the low multipole \enquote{Sachs-Wolfe} plateau (related to the redshift of dark radiation domination, giving a dependence on $\Omega_m$), and the diffusion damping of higher-order peaks. Only in their combination, and within a model, especially with the radiation density scale inferred by the, exquisitely measured by COBE/FIRAS, photon temperature, can these additional effects eliminate the various degeneracy directions and result in a combined Hubble parameter inference. It should be stressed at this point, that to obtain the Hubble parameter from the CMB a variety of model assumptions about acoustic oscillations and expansion rates at high and low redshift have to be combined.

\subsection{A note on RSD}\label{ssec:RSD}
The galaxy clustering 3D maps  can also provide  a direct measurement of the linear growth rate, $f$, through the redshift space distortions (RSD, \cite{kaiser_clustering_1987}). RSD are due to peculiar velocities sourced by inhomogeneities and distort the observed clustering introducing anisotropies with respect to the line of sight. In practice, the quantity best constrained by RSD is the combination $f\sigma_8$. In general relativity the growth history is dictated by the expansion history, thus in principle RSD can offer a powerful consistency check, which will be exploited by future surveys. At present, this test is mostly qualitative, as RSD information does not improve the $H_0$ determination. 

\end{document}

%% file: ads_macros.tex
\let\jnlstyle=\rm
\def\ref@jnl#1{{\jnlstyle#1}}

\def\aj{\ref@jnl{AJ}}                   
\def\actaa{\ref@jnl{Acta Astron.}}      
\def\araa{\ref@jnl{ARA\&A}}             
\def\apj{\ref@jnl{ApJ}}                 
\def\apjl{\ref@jnl{ApJ}}                
\def\apjs{\ref@jnl{ApJS}}               
\def\ao{\ref@jnl{Appl.~Opt.}}           
\def\apss{\ref@jnl{Ap\&SS}}             
\def\aap{\ref@jnl{A\&A}}                
\def\aapr{\ref@jnl{A\&A~Rev.}}          
\def\aaps{\ref@jnl{A\&AS}}              
\def\azh{\ref@jnl{AZh}}                 
\def\baas{\ref@jnl{BAAS}}               
\def\bac{\ref@jnl{Bull. astr. Inst. Czechosl.}}
\def\caa{\ref@jnl{Chinese Astron. Astrophys.}}
\def\cjaa{\ref@jnl{Chinese J. Astron. Astrophys.}}
\def\icarus{\ref@jnl{Icarus}}           
\def\jcap{\ref@jnl{J. Cosmology Astropart. Phys.}}
\def\jrasc{\ref@jnl{JRASC}}             
\def\memras{\ref@jnl{MmRAS}}            
\def\mnras{\ref@jnl{MNRAS}}             
\def\na{\ref@jnl{New A}}                
\def\nar{\ref@jnl{New A Rev.}}          
\def\pra{\ref@jnl{Phys.~Rev.~A}}        
\def\prb{\ref@jnl{Phys.~Rev.~B}}        
\def\prc{\ref@jnl{Phys.~Rev.~C}}        
\def\prd{\ref@jnl{Phys.~Rev.~D}}        
\def\pre{\ref@jnl{Phys.~Rev.~E}}        
\def\prl{\ref@jnl{Phys.~Rev.~Lett.}}    
\def\pasa{\ref@jnl{PASA}}               
\def\pasp{\ref@jnl{PASP}}               
\def\pasj{\ref@jnl{PASJ}}               
\def\rmxaa{\ref@jnl{Rev. Mexicana Astron. Astrofis.}}%
\def\qjras{\ref@jnl{QJRAS}}             
\def\skytel{\ref@jnl{S\&T}}             
\def\solphys{\ref@jnl{Sol.~Phys.}}      
\def\sovast{\ref@jnl{Soviet~Ast.}}      
\def\ssr{\ref@jnl{Space~Sci.~Rev.}}     
\def\zap{\ref@jnl{ZAp}}                 
\def\nat{\ref@jnl{Nature}}              
\def\iaucirc{\ref@jnl{IAU~Circ.}}       
\def\aplett{\ref@jnl{Astrophys.~Lett.}} 
\def\apspr{\ref@jnl{Astrophys.~Space~Phys.~Res.}}
\def\bain{\ref@jnl{Bull.~Astron.~Inst.~Netherlands}} 
\def\fcp{\ref@jnl{Fund.~Cosmic~Phys.}}  
\def\gca{\ref@jnl{Geochim.~Cosmochim.~Acta}}   
\def\grl{\ref@jnl{Geophys.~Res.~Lett.}} 
\def\jcp{\ref@jnl{J.~Chem.~Phys.}}      
\def\jgr{\ref@jnl{J.~Geophys.~Res.}}    
\def\jqsrt{\ref@jnl{J.~Quant.~Spec.~Radiat.~Transf.}}
\def\memsai{\ref@jnl{Mem.~Soc.~Astron.~Italiana}}
\def\nphysa{\ref@jnl{Nucl.~Phys.~A}}   
\def\physrep{\ref@jnl{Phys.~Rep.}}   
\def\physscr{\ref@jnl{Phys.~Scr}}   
\def\planss{\ref@jnl{Planet.~Space~Sci.}}   
\def\procspie{\ref@jnl{Proc.~SPIE}}   